\documentclass[reprint,amsmath,amssymb, aps, pra]{revtex4-1}

\usepackage{graphicx}
\usepackage{natbib}
\usepackage{color}
\definecolor{mygrey}{gray}{0.35}
\definecolor{myblue}{rgb}{0.2,0.2,0.8}
\definecolor{myzard}{cmyk}{0,0,0.05,0}
\definecolor{mywhite}{rgb}{1,1,1}
\definecolor{myred}{rgb}{1,0.,0.3}
\usepackage[colorlinks=true,citecolor=myblue,linkcolor=myred]{hyperref}
\begin{document}
\title{Rabi Lattice models with discrete gauge symmetry: phase diagram and implementation in trapped ion quantum simulators}
\author{Pedro Nevado}
\email{P.Nevado-Serrano@sussex.ac.uk}
\author{Diego Porras}
\email{D.Porras@sussex.ac.uk}
\affiliation{%
Department of Physics and Astronomy, University of Sussex, Falmer, Brighton BN1 9QH, UK
}
\date{\today}
\begin{abstract}
We study a spin-boson chain that exhibits a local $\mathbb{Z}_2$ symmetry. We investigate the quantum phase diagram of the model by means of perturbation theory, mean-field theory and the Density Matrix Renormalization Group method. Our calculations show the existence of a first-order phase transition in the region where the boson quantum dynamics is slow compared to the spin-spin interactions. Our model can be implemented with trapped ion quantum simulators, leading to a realization of minimal models showing local gauge invariance and first-order phase transitions.
\end{abstract}
\pacs{Valid PACS appear here}
\maketitle

\section{Introduction}
Analogical quantum simulators with many-body optical setups offer us the possibility  to replicate the physics of condensed matter systems, and also to engineer novel exotic quantum phases 
\cite{Cirac12natphys}. 
In particular, trapped ions 
\cite{Friedenauer08natphys,Schneider12rpp,Blatt12natphys} and superconducting circuits 
\cite{Houck12natphys} are ideally suited to implement lattice models of spins coupled to bosons with a wide control of spin-boson and spin-spin interactions. The resulting family of models that can be directly simulated in those setups include cooperative Jahn-Teller and Rabi Lattice Hamiltonians \cite{Porras12prl,Tureci12prl,Koch13adp,Kurcz14prl}. The physical implementation of those models lead us to the exciting possibility to study complex quantum phases governed by the interplay between magnetic and vibronic or photonic degrees of freedom. Furthermore, the fabrication of arrays of ion microtraps open up a new perspective to control lattice geometry and particle interactions \cite{Home09sci,Schmied09prl,Kumph11njp,Hensinger14natcom}.

In this work, we introduce a Rabi Lattice model that shows a local (gauge) discrete invariance, something that takes this model out of the universality classes that we typically find in strongly correlated spin-boson lattice systems. 
The implementation of lattice gauge theories with trapped ions and superconducting circuits has been proposed in recent works 
\cite{Marcos13prl,Hauke13prx}. Here we take a different approach to find out the simplest minimal Rabi Lattice model that shows local gauge invariance and can be implemented in many-body quantum optical setups. In fact, spin-boson couplings can lead in a natural way to the appearance of a discrete local gauge invariance. Consider for example the case of an Ising model, with a Hamiltonian of the form 
\begin{equation}
H_{\rm I} = \sum_j h_j \sigma^x_j - J \sum_{j}\sigma^z_j \sigma^z_{j+1}.
\label{q.Ising}
\end{equation}

Local discrete gauge invariance may appear when we replace the local field $h_j$ by a quantum variable, for example, the position operator of a local bosonic field,
\begin{equation}
h_j \to g (a_j + a_j^\dagger).
\nonumber
\end{equation}

After this substitution, we get an Ising spin model where the transverse field is a variable with quantum dynamics of its own. This Hamiltonian possess a discrete local symmetry, since it is invariant under a set of local transformations defined at each site $j$, $\sigma^x_j \to - \sigma^x_j$, $a_j \to - a_j$. The model turns out to be a Rabi Lattice, where different sites are coupled by an Ising interaction between spins.

This work is organized as follows. Motivated by the discussion above, in Section II, we introduce the Ising-Rabi Lattice model and its symmetry properties. In Section III we discuss the ground state of the model in some limiting cases by using perturbation theory. In Section IV we present two variational ans\"atze to approximately find the ground-state of our model: a Born-Oppenheimer approximation, valid in the limit in which bosonic degrees of freedom are slow compared to the spin dynamics, and a Silbey-Harris approach valid in the limit of fast bosonic modes. 
Those approximations predict a first-order phase transition between a pure ferromagnetic Ising phase and a dressed ferromagnetic phase of displaced bosons. In section V we present numerical results obtained with the Density Matrix Renormalization Group (DMRG) method that confirm the validity of the Born-Oppenheimer approximation and the existence of a first-order phase transition. Section VI presents a proposal to implement our model with trapped ions in arrays of microtraps. Finally we present our conclusions in Section VII.

\section{Ising-Rabi Lattice Hamiltonian}

We introduce the one-dimensional Ising-Rabi Lattice Hamiltonian. Our system consists of $N$ spins arranged in a 1D chain interacting via a nearest neighbours exchange Ising coupling term of strength $J$ (we will assume $J>0$ for definiteness on the following, but the results are equivalent if $J<0$). Spins are coupled to local bosonic modes of energy $\delta>0$ by an on-site spin-dependent force of magnitude $g$,
\begin{equation}
H_{\rm IR} = \delta \sum_{j=1}^N a^{\dagger}_j a_j + 
g \sum_{j=1}^{N} \sigma^x_j (a^{\dagger}_j + a_j) 
- J \sum_{j=1}^{N-1} \sigma^z_j \sigma^z_{j+1}.
\label{hamiltonian}
\end{equation}

This model possesses a {\it gauge}, (i.e., that acts locally, independently on every site) $\mathbb{Z}_2$ symmetry, since it is invariant with respect to the transformation prescribed by
\begin{equation}
\mathcal{P}_{\rm gauge}^{(j)}=e^{i\pi \left(a_j^{\dagger} a_j + \frac{\sigma^z_j}{2}\right)},\, \text{so} \left[H,\mathcal{P}_{\rm gauge}^{(j)}\right]=0\quad \forall j,
\end{equation}
that transforms operators $a_j \rightarrow -a_j$, $ \sigma^x_j \rightarrow -\sigma^x_j$, but leaves invariant the Ising coupling term, since $\sigma^z_j \rightarrow \sigma^z_j$.  
We expect that this discrete and local symmetry cannot be spontaneously broken in the ground state of the Hamiltonian, a result that is expected to generally apply to any local symmetry, and is referred to as Elitzur's Theorem \cite{kogut79rmp}. 
Accordingly, expectation values $\langle  a_j \rangle_{\rm GS} = 0$ and $\langle \sigma^x_j \rangle_{\rm GS} = 0$ in the whole phase diagram of the model.

The Hamiltonian (\ref{hamiltonian}) also possesses a global $\mathbb{Z}_2$ symmetry related to the transformation  $\sigma^z_j \rightarrow -\sigma^z_j,\,\forall j$, whose representation in the current space of states is given by the unitary operator
\begin{equation}
\mathcal{P}=e^{i\pi \mathcal{N}},\text{with \,} \mathcal{N} = \sum_{j=1}^{N} \frac{\sigma^x_j}{2}.
\end{equation}

Since $\left[H_{\rm IR},\mathcal{P}\right]=0$, the ground state (GS) should fulfil $\langle \sigma^z_j \rangle_{\rm GS} = 0$, unless degeneracy occurs. This global symmetry is actually also present in the quantum Ising model, where it is spontaneously broken in the ferromagnetic phase, such that 
$\langle \sigma^z_j \rangle_{\rm GS} \neq 0$ in the thermodynamical limit. 
However, in a finite size quantum Ising chain a linear superposition of ferromagnetic states can form the ground state, leading to $\langle \sigma^z_j \rangle_{\rm GS} = 0$ for finite $N$. Below we show that a remarkable feature of the Ising-Rabi Lattice Hamiltonian is the existence of symmetry breaking of the global parity symmetry for finite values of $N$.

\section{Asymptotic limits of the Ising-Rabi Lattice}
\subsection{Ferromagnetic phase}
We discuss the limit $\delta , J \gg g$, that leads to a {\it ferromagnetic} (F) phase. 
We define $H^0_{F}$ by considering the limit $g = 0$ of the IR model, 
\begin{equation}
H_{\rm F}^0 = \delta \sum_{j=1}^{N} a_j^{\dagger} a_j - J \sum_{j=1}^{N-1}  \sigma_j^z \sigma_{j+1}^z.
\end{equation}

The ground states of $H_{\rm F}^0$ consist of the boson vacuum and one of the possible ferromagnetic orders (cf. Ising model \cite{ChakrabartiBook}). We will refer to these states as 
\begin{equation}
\begin{aligned}
| \phi_{{\rm F}, \uparrow} \rangle = 
| 0 \rangle_{\rm b}  \bigotimes_{j=1}^N | \uparrow_{z} \rangle_j , \\
| \phi_{{\rm F}, \downarrow} \rangle = 
| 0 \rangle_{\rm b}  \bigotimes_{j=1}^N | \downarrow_{z} \rangle_j , 
\end{aligned}
\label{ferro.states}
\end{equation}
to make a explicit choice of basis in the two-fold degenerate manifold. To study the stability of the ferromagnetic phase, we introduce the spin boson coupling as a perturbation,
\begin{eqnarray}
H_{\rm F}' = g \sum_{j=1}^{N} \sigma_j^x (a_j^{\dagger} + a_j),
\end{eqnarray}
and consider its effect upon the degenerate manifold of ground states. By applying degenerate perturbation theory, we find that $H_{\rm F}'$ does not lift the degeneracy {\it even at finite $N$} (see Appendix (\ref{App:A}) for details). This situation is in clear contrast with the quantum Ising model, where the degeneracy is lifted in the ferromagnetic phase by an energy gap scaling like $\propto h^{N}$ \cite{Ivanov13pra}, with $h$ the value of transverse field in Eq. (\ref{q.Ising}).

By using perturbation theory we calculate the energy of any of the degenerate ferromagnetic ground states, including the leading corrections induced by the spin-boson coupling,
\begin{equation}
E_{\rm F} \simeq -J(N-1) -g^2\left[\frac{N-2}{\delta +4 J}+\frac{2}{\delta + 2 J}\right].
\label{energy.ferro}
\end{equation}

Perturbation theory also predicts that states (\ref{ferro.states}) are a good approximation to the ground state of $H_{\rm IR}$ as long as $g\ll \delta +4 J$.

\subsection{Dressed ferromagnetic phase}
\label{section.dressed.ferro}
We consider now the limit $g, \delta \gg J$, where the Ising interaction is small compared to the spin-boson coupling and the boson energies. 
Here we can perform a boson-displacement unitary transformation \cite{Porras04prl} in (\ref{hamiltonian}), considering as well the rotation $x \leftrightarrow z$: $H_{\rm IR}\rightarrow \bar{H}_{\rm IR} = U \mathcal{R}_{xz} H_{\rm IR} \mathcal{R}_{xz}^{\dagger} U^{\dagger}$, with $\mathcal{R}_{xz} = 1/2^{N/2} \bigotimes_{j=1}^N(\sigma_j^x + \sigma_j^z)$, and
\begin{equation}
U = \bigotimes_{j=1}^N e^{S_j},\quad S_j 
= \frac{g}{\delta}  \sigma_j^z (a_j^{\dagger} - a_j),
\label{polaron}
\end{equation}
so that the IR Hamiltonian reads $\bar{H}_{\rm IR}= \bar{H}_{\rm DF}^0+\bar{H}_{\rm DF}'$, with $\bar{H}_{\rm DF}^0 = \delta \sum_{j=1}^N a^{\dagger}_j a_j - N g^2/\delta$, and
\begin{equation}
\bar{H}_{\rm DF}'=-  J \sum_{j=1}^{N-1} (\bar{\sigma}^+_j+\bar{\sigma}^-_j)(\bar{\sigma}^+_{j+1}+\bar{\sigma}^-_{j+1}).
\label{displaced.hamiltonian}
\end{equation}

We have defined operators $\bar{\sigma}^{\pm}_j = e^{\pm 2 S_j} \sigma^{\pm}_j$, with $S_j$ according to equation (\ref{polaron}). The ground states of $\bar{H}_{\rm DF}^0$ consist of the vacuum of the bosons in the displaced basis, and for any spin configuration. However, this degeneracy is removed considering the action of the perturbation upon these states, 
%
\begin{equation}
\,_{\rm b} \langle 0| \bar{H}_{\rm DF}'|0\rangle_{\rm b} = -t 
e^{-4\alpha^2} \sum_{j=1}^{N-1} \sigma^x_j \sigma^x_{j+1},\ \ \ \alpha = \frac{g}{\delta},
\label{dressed.ferro}
\end{equation}
which shows that the ground states of $\bar{H}_{\rm IR}$ when $t\to 0$ are just the two ferromagnetic states in the $x$ direction. Therefore, transforming these states back to the original basis, we find the two degenerate ground states of (\ref{hamiltonian}),
\begin{equation}
| \phi_{{\rm DF},\pm} \rangle = \frac{1}{2^{N/2}} \bigotimes_{j=1}^N (|-\alpha,\uparrow_{x}\rangle_j \pm|\alpha,\downarrow_{x}\rangle_j),
\label{dressed_states}
\end{equation}
and we will refer to them as {\it dressed-ferromagnetic} (DF) states. The energy of these states, together with the leading order correction induced by the dressing boson operators in Hamiltonian (\ref{displaced.hamiltonian}) is given by
\begin{equation}
E_{\rm DF}  \simeq  -\frac{Ng^2}{\delta} -  J(N-1) e^{-4\alpha^2} -(N-1)\frac{J^2}{\delta} P(\alpha),
\label{energy.dressed_ferro}
\end{equation}
where we have defined
\begin{equation}
P(\alpha)=\sum_{p=1}^{\infty} \frac{1}{p} \frac{e^{-8\alpha^2}  (8\alpha^2)^p}{p!}.
\label{P.function}
\end{equation}

Note that $P(\alpha) \to 0$ if $\alpha \to 0$, and $P(\alpha) \to (8 \alpha^2)^{-1}$ if $\alpha \to \infty$. The DF state is perturbed by a correction $(J/\delta)\, P(g/\delta)$. The latter is negligible if $\delta \gg J$ in the limit $g \ll \delta$, and if $g^2 \gg J \delta$ in the limit $g \gg \delta$.

\subsection{Qualitative discussion of the quantum phase diagram}

The previous considerations allow us to make a conjecture about the phase diagram. We distinguish two cases: 

(i) $\delta \ll J$. In this limit, condition $g \ll J$ ensures that the F states (\ref{ferro.states}) are possible ground states of $H_{\rm IR}$. Following the discussion below Eq. (\ref{energy.dressed_ferro}), the DF states are possible ground states if $g \gg \sqrt{J \delta}$. In the interval $\sqrt{J \delta} < g < J$, the domain of F and DF solutions overlap, and we expect a crossover between those energy levels. Comparing the F and DF energy, we find that crossover at $g:=g_{\rm c} >\sqrt{J\delta}$, where we expect the appearance of a first order F-DF transition.

(ii) $\delta \gg J$. Here, F states are valid ground states if $g \ll \delta$, whereas DF states are valid ground states for any value of $g$, as follows from the discussion below Eq. (\ref{energy.dressed_ferro}). In the interval $g \ll \delta$, F and DF solutions overlap, however, here the DF state continuously converges to the F state. Thus we expect a continuous transition from the DF to the F solution.

\begin{figure}[h!]
	\includegraphics[width=3.2in]{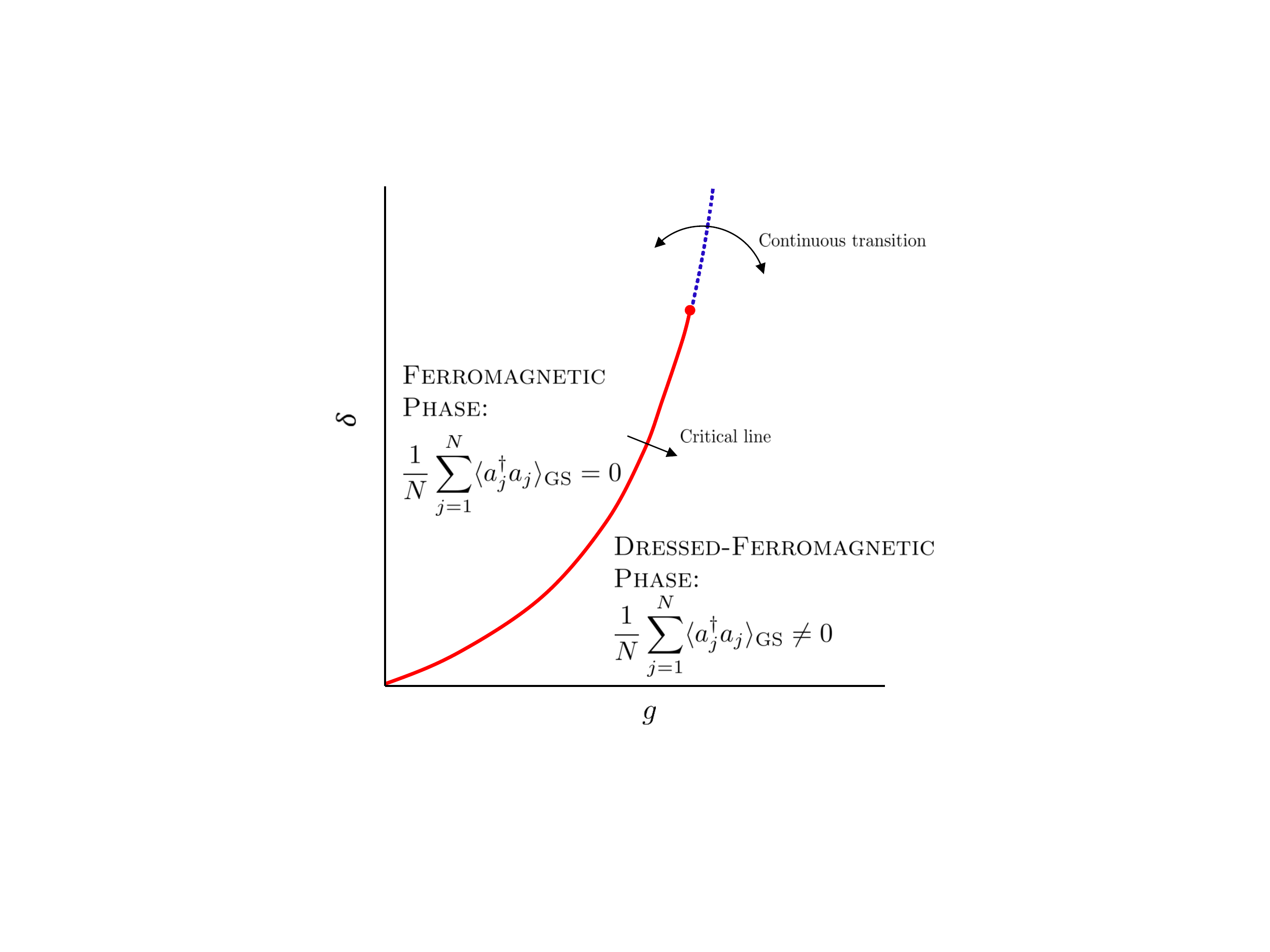}
	\caption{Scheme of the phase map depicting the disappearance (at the dot) of the discontinuous jump in the number of bosons along the critical line (solid) for a given value of $\delta,\,g$. The dashed line represents no boundary but a continuous transition from the ferromagnetic to the dressed-ferromagnetic phase.}
	\label{phase_map}
\end{figure}

Putting together all previous arguments, we expect that $H_{\rm IR}$ presents a first order quantum phase transition along the critical line $g_{\rm c} (\delta, J)$, featuring a jump from the F to the DF ground states in the regime of low boson energies $\delta \to 0$. This is in clear contrast with the quantum Ising chain with a transverse field, where there is no coexistence of the ferro- and paramagnetic phases at neither side of the (second order) phase transition. In the $H_{\rm IR}$, however, there is a coexistence of the phases already addressed if $\sqrt{\delta J} \ll g\ll J$. Furthermore, this last set of inequalities cannot be longer fulfilled if $\delta \gg J$, and therefore the discontinuous behaviour is bound to disappear for a given $\delta \sim J$. We have summarized these considerations in Fig. \ref{phase_map}, where we choose as order parameter the average boson number
\begin{equation}
n = \frac{1}{N} \sum_{j=1}^N \langle a_j^{\dagger} a_j \rangle
\end{equation}
to capture the sudden change from the boson vacuum state (F phase) to a displaced state (DF phase).

Below we consider two different mean-field descriptions that give some physical insight on the phases and the transition of the problem. In addition, they will be validated afterwards by a exact numerical DMRG diagonalization.

\section{Variational methods}
\subsection{Born-Oppenheimer approximation ($\delta \ll J$)}
The classical limit of the model is attained in the regime of very high number of bosonic excitations, which is expected at $\delta \to 0$. In this limit, ladder operators can be treated as classical variables, $a_j \to \alpha_j \in \mathbb{C}$, and $H_{\rm IR}$ is reduced to a spin Hamiltonian,
\begin{eqnarray}
&& H_{\rm BO} = \nonumber \\
&& \ \ \delta \sum_{j=1}^N |\alpha_j|^2 + g\sum_{j=1}^{N} \sigma^x_j (\alpha^*_j + \alpha_j) - J \sum_{j=1}^{N-1} \sigma^z_j \sigma^z_{j+1}. 
\label{spin.hamiltonian}
\end{eqnarray}

Eq. (\ref{spin.hamiltonian}) describes an Ising chain in a transverse field, for which an exact ground state $|\Psi_{\rm I}(\alpha_j)\rangle$ can be found \cite{ChakrabartiBook}. Without loss of generality we can assume $\alpha_j$ to be real. We devise a variational ansatz by calculating the mean value of $H_{\rm BO}$, whose ground state energy can be written as
$E_{\rm BO}(\{\alpha_j\}) = \delta \sum_{j=1}^N \alpha_j^2 + E_{\rm I,0}(\{\alpha_j\})$, 
where $E_{\rm I,0}(\{\alpha_j\})$ is the ground state energy of the quantum Ising chain (\ref{q.Ising}) with transverse fields $h_j = 2 g \alpha_j$ and interaction strength $J$. The corresponding variational wavefunction is
\begin{equation}
|\Psi_{\rm BO}\rangle = |\Psi_{\rm I}(\alpha_j)\rangle \bigotimes_{j=1}^N|\alpha_j\rangle.
\label{BO}
\end{equation}
This method is a self-consistent approach that resembles the Born-Oppenheimer approximation in Molecular Physics \cite{BaymBook}. In that context, the degrees of freedom of the positions of the nuclei enter the electronic Hamiltonian as parameters in the same way the boson amplitudes appear in the spin Hamiltonian (\ref{spin.hamiltonian}). We notice that, due to the underlying gauge symmetry in the $H_{\rm IR}$ Hamiltonian, a variational solution of the form (\ref{BO}) can be transformed into a solution with the same energy if we change locally the sign of the displacement $\alpha_j$, and simultaneously transform $\sigma^x_j \to - \sigma^x_j$. There are thus $2^N$ possible solutions, given by the values $\alpha_j = s_j |\alpha_j|$, with $s_j = \pm 1$.

In order to make best use of the analytical results for the solution of (\ref{spin.hamiltonian}), we assume $N \to \infty$ and $\alpha_j \to -\alpha$ (in the thermodynamic limit the system is homogeneous, whereas the minus sign is chosen for analytical convenience), so the energy $E_{\rm BO}$ of the ground state of $H_{\rm BO}$ is
\begin{equation}
\frac{E_{\rm BO}}{N} = \delta \alpha^2 -2\alpha g \frac{2}{\pi}(1+\lambda) E\left[\frac{4\lambda}{(1+\lambda)^2}\right], \lambda=\frac{J}{2\alpha g},
\label{energy_constraint}
\end{equation}
where $E$ is the complete elliptic integral of the second kind. We are interested in the value of the parameter $\alpha$ for which the energy attains a minimum; we will refer to this point as $\alpha_0$, and its value together with the exact solution of the spin problem will define the mean-field ground state. 

\begin{figure}[h]
	\begin{flushleft}
	\includegraphics[width=3.4in]{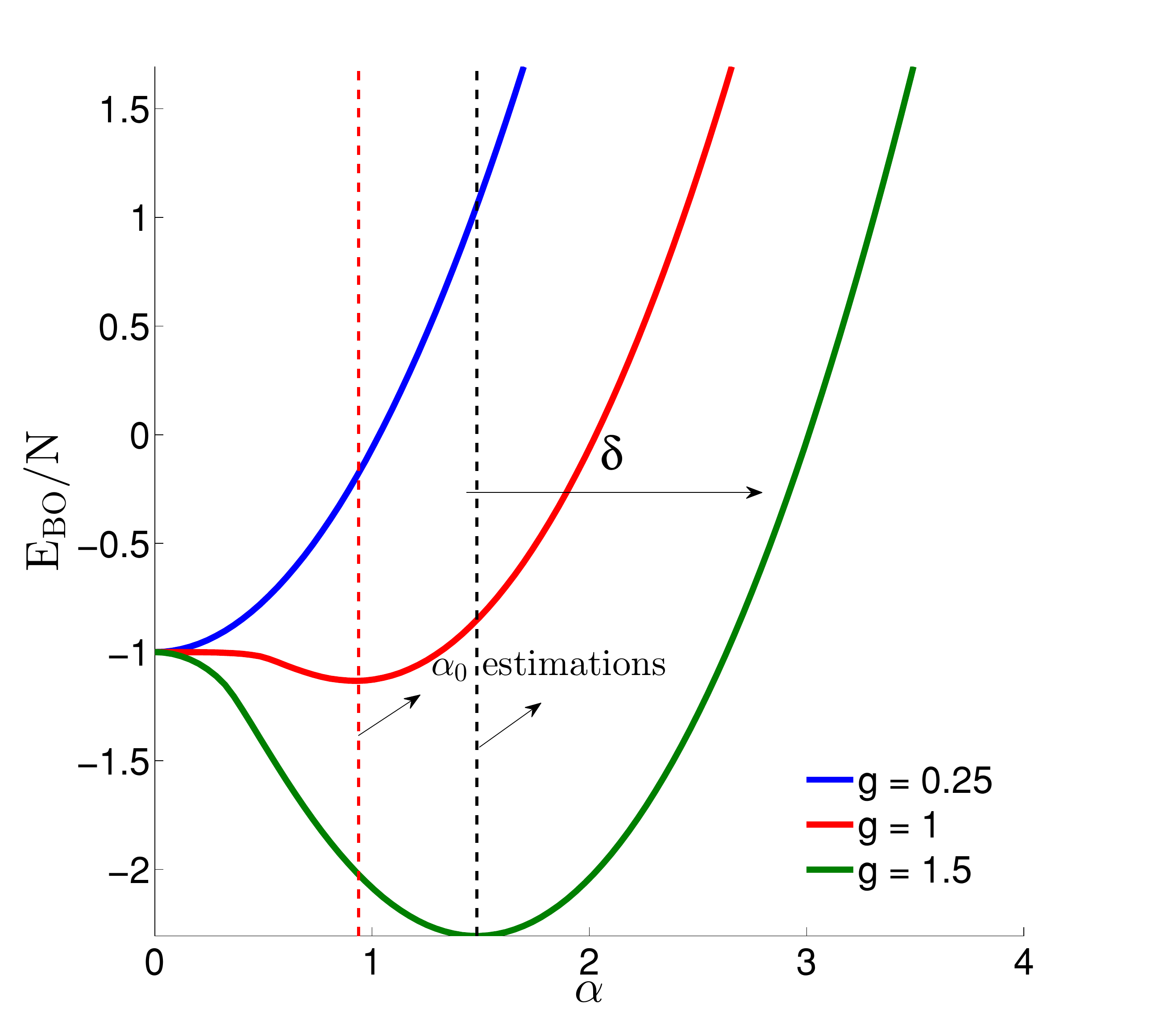}
	\caption{Mean field energy (\ref{energy_constraint}) for $\delta=J=1$ as a function of the parameter $\alpha$ and different values of the spin-phonon coupling $g$. Note that close to the origin there is a curvature change for a given $g \geq g_c$. This point marks the criticality condition.}
	\label{mean_field}
	\end{flushleft}
\end{figure}

A quick inspection of (\ref{energy_constraint}) reveals that as a function of $\alpha$, the energy is minimum exclusively at the origin unless there are values of $J,g$ that shift its position to a finite value $\alpha \neq 0$ (see Fig. \ref{mean_field}). Bearing this in mind we carry out the Taylor expansion of the energy around $\alpha=0$, that leads to
\begin{equation}
\frac{E_{\rm BO}}{N} = -J + (\delta - \frac{g^2}{J})\alpha^2 + O(\alpha^4),
\end{equation}
and predicts a minimum for $\alpha_0\neq 0$ ($\alpha_0=0$) whenever $g^2 \geq \delta J:=g_{\rm c}^2$ ($g<g_{\rm c}$). We interpret that the system undergoes a phase transition at that point: below the critical line bosonic excitations are inhibited ($\alpha_0=0$), and the spins point in the ${+z}$ or ${-z}$ directions; above $g_{\rm c}$, the ground state changes abruptly to allow an arbitrary number of bosons, whereas the spins point in the direction determined by the Ising ground state for a transverse field of magnitude $2\alpha_0 g$. Furthermore, in this latter regime we can estimate the value of $\alpha_0$ assuming $g \gg J$ --for fixed $\alpha,\delta$--, which gives
\begin{equation}
\alpha_0=\frac{g}{\delta}\left(1-\frac{J^2\delta^2}{16g^4}\right).
\label{number.of.bosons.bo}
\end{equation}

From the previous discussion we can extract the order parameter $n = \alpha_0^2$, that we shall compare with the DMRG results to assess the validity of the previous approximations.

The Born-Oppenheimer solution converges to the DF states in the limit $g \gg \delta$. In this limit one can easily show that the optimal values are $|\alpha_j| = g/\delta$. We can restore the $\mathbb{Z}_2$ gauge symmetry by considering a symmetric superposition,
\begin{equation}
|\Psi_{\rm BO}^{\rm sym}\rangle = \frac{1}{2^{N/2}} \sum_{\substack{s_1,\ldots,s_N\\ s_j=\pm 1}} |\Psi_{\rm I} \left(s_j \frac{g}{\delta} \right) \rangle 
\bigotimes_{j = 1}^N | s_j \frac{g}{\delta} \rangle.
\end{equation}
such that we recover the solution $|\phi_{\rm DF,+}\rangle$. The solution $|\phi_{\rm DF,-}\rangle$ would correspond to the antisymmetric linear combination of the former states.

\subsection{Silbey-Harris-type ansatz ($\delta \gg J$)}

In order to investigate the continuous transition regime mentioned in Fig. \ref{phase_map}, we are going to consider a displaced trial wave-function whose distance away from the origin in phase space is no longer fixed, rather the variational parameter \cite{Silbey84jourchemphys}. This approach has been recently shown to yield an accurate description of the quantum phase diagram in Rabi Lattice models \cite{Kurcz14prl,Kurz14arX}.

Specifically, we take the IR Hamiltonian in the rotated basis $x \leftrightarrow z$, and compute its energy upon the wave-function
%
\begin{equation}
|\Psi_{\rm SH}\rangle = e^{-S(\eta)} |0\rangle_{\rm b} \bigotimes_{j=1}^N\left|\uparrow_x\right\rangle_j, S(\eta)=  \eta \frac{g}{\delta}\sum_{j=1}^N \sigma^z_j (a^{\dagger}-a),
\end{equation}
where the parameter $\eta$ continuously interpolates the displaced solution between $0$ and $g/\delta$ for fixed values of these. The Silbey-Harris energy reads
\begin{equation}
E_{\rm SH} (\eta)= N\frac{g^2}{\delta} (\eta^2-2\eta)- J(N-1)  e^{-4 \eta^2 \left(\frac{g}{\delta}\right)^2},
\label{silbey.harris.condition}
\end{equation}
which along with the condition $d E_{\rm SH}/d\eta = 0$ for a given $\eta = \eta_0$, leads to the optimal value for the order parameter  $n = (\eta_0 g /\delta)^2$ within this framework.

\begin{figure}[h!]
	\begin{flushleft}
	\includegraphics[width=3.2in]{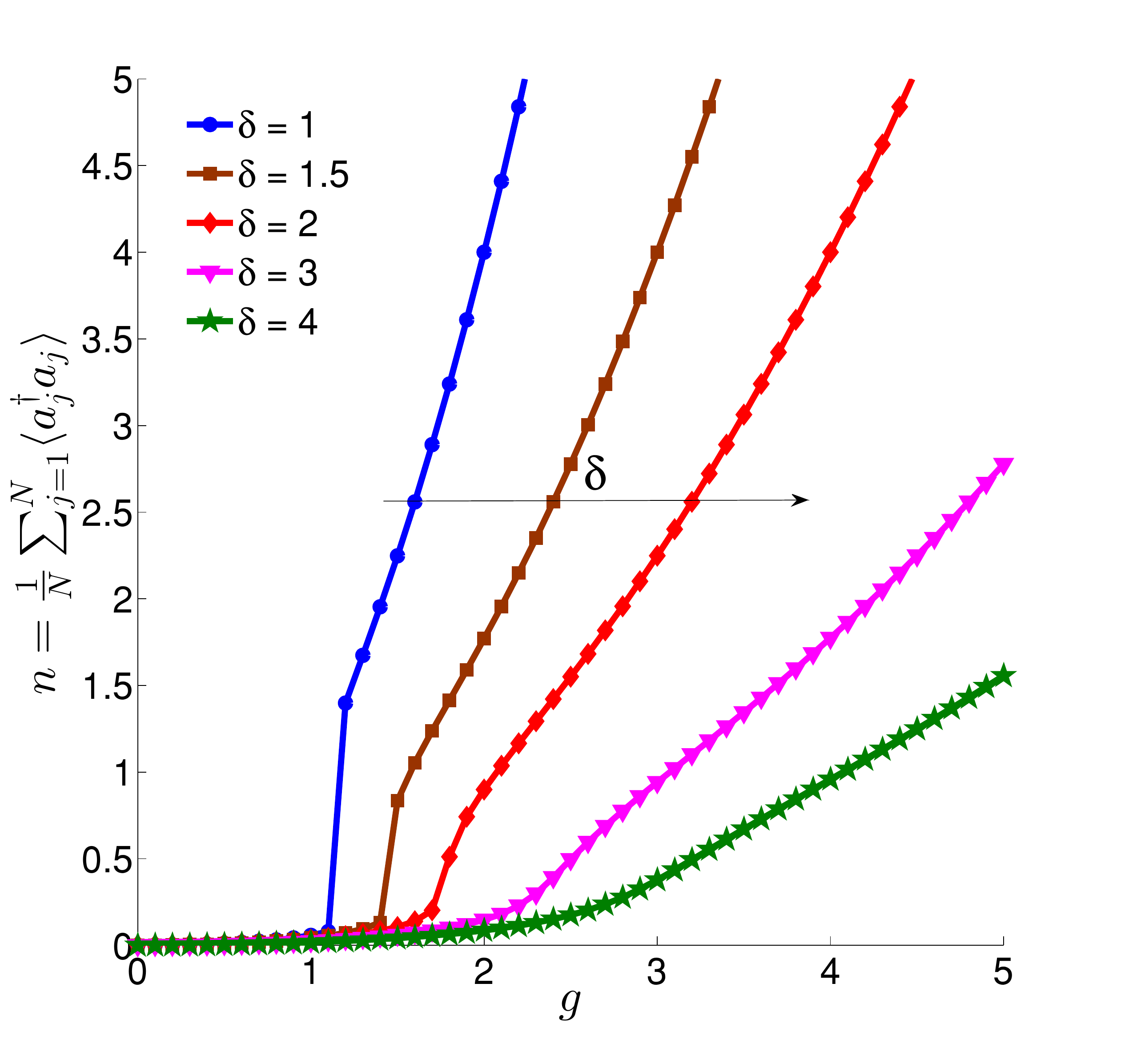}
    \caption{Silbey-Harris mean boson number $n$ for different values of $\delta$, $J=1$ and $N=50$ sites.}
    \label{silbey.harris.vs.delta}
	\end{flushleft}
\end{figure}

This ansatz resembles the exact IR Hamiltonian solution if $\delta \gg J$, because in that case the ground state is one of the dressed-ferromagnetic eigenvectors (cf. section \ref{section.dressed.ferro}). However, it turns out that it also predicts a first order phase transition when extrapolated to the $\delta\ll J$ regime. This supports the fact that $H_{\rm IR}$ exhibits a sudden ground state change in this latter case (see Fig. \ref{silbey.harris.vs.delta}).

We present the predictions of the Silbey-Harris solution, focusing on the fact that the discontinuity of $n$ at the transition disappears between the regimes $\delta \gg J$ and $\delta \ll J$.

\section{DMRG results}

In this section we present quasi-exact numerical calculation of the ground state properties of the IR Hamiltonian for a chain of $N=50$ spins, obtained by means of the DMRG algorithm \cite{White93prb}. Some remarks are in order before proceeding to the results. 
First, we have to introduce a cut-off, $N_{\rm c}$, in the maximum Fock state of local bosonic modes in the DMRG algorithm. This imposes some limitations in the description of the DF phase in the limit $\delta \ll g$. Here, due to the low energy cost of bosonic excitations, the ground state wavefunction has non-negligible projections upon many different occupation states. Thus, an accurate description may require high values of $N_{\rm c}$ that are beyond our computational capabilities. In this work we use $N_{\rm c} = 10$, and present exclusively DMRG results fulfilling 
$2 n \leq N_{\rm c}$. Second, we stress that finite-size effects in our calculations lead to a smearing of discontinuities at the first order phase transition, which strictly speaking takes place in the thermodynamic limit only.

\begin{figure}[h!]
	\begin{flushleft}
	\includegraphics[width=3.5in]{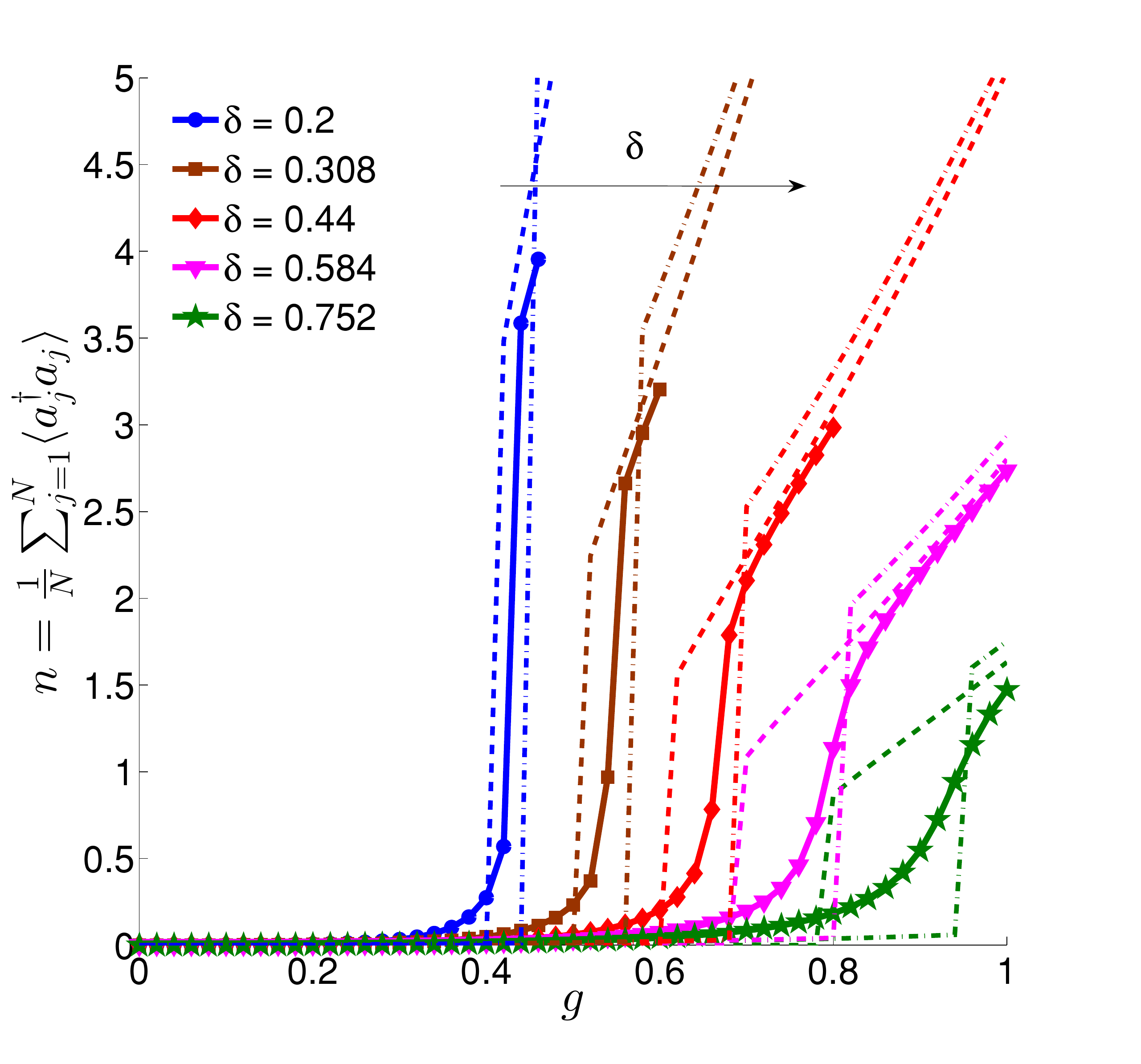}
    \caption{Mean boson number prediction for Born-Oppenheimer (dashed lines), Silbey-Harris-type ansatz (dashed-dotted lines) and DMRG (solid lines) of a $N=50$ sites chain, and $J=1$. For the DMRG method we set a renormalization dimension $D = 10$, on-site boson cut-off $N_{\rm c} = 10$ and local dimension $d = 2 \cdot  N_{\rm c}$ (we follow the notation of \cite{Schollwoeck11aop}).}
    \label{Ncomparisons}
	\end{flushleft}
\end{figure}

\begin{figure}[h!]
	\begin{flushleft}
	\includegraphics[width=3.8in]{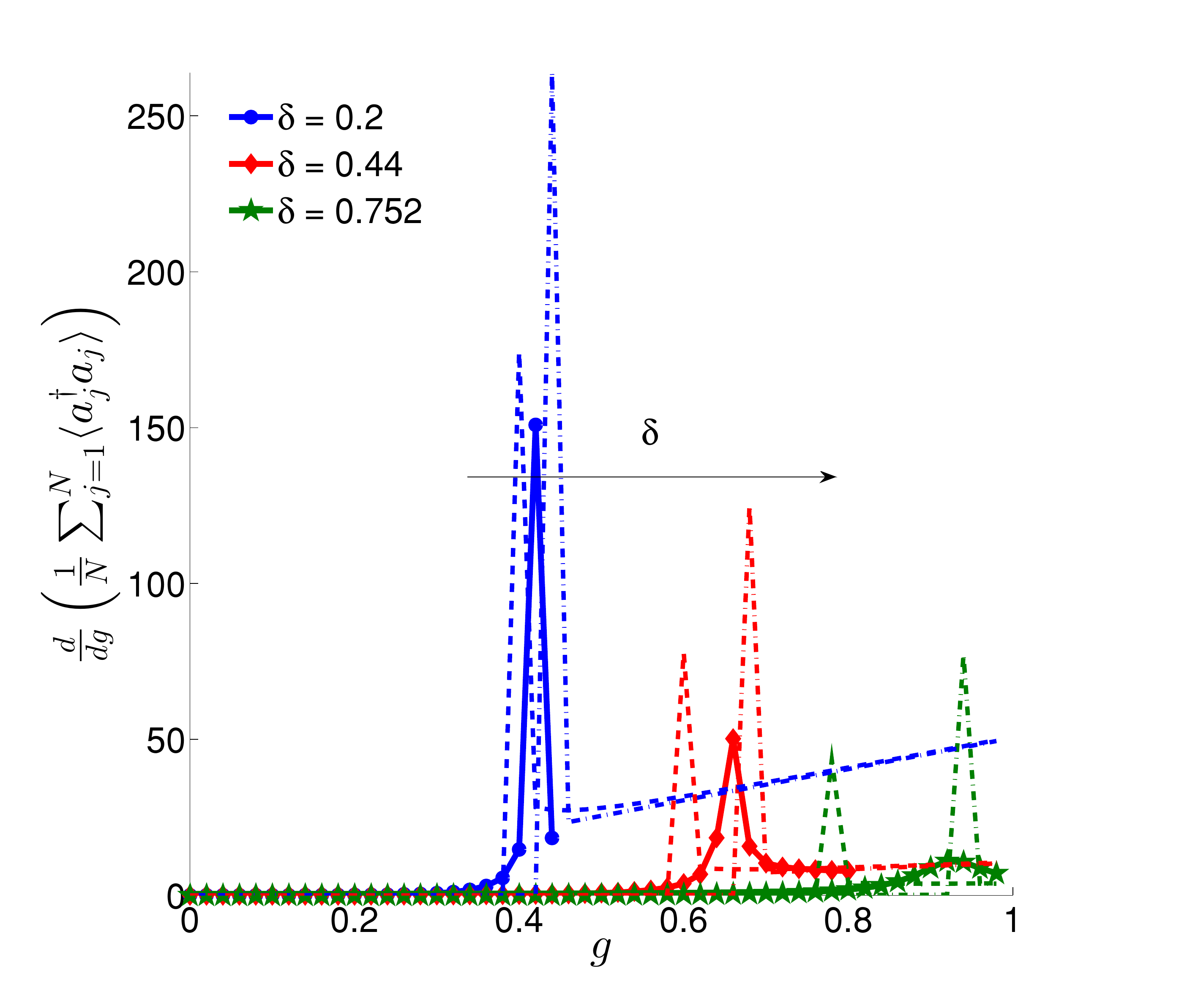}
    \caption{Derivatives of the average boson number for different values of $\delta$ ($N=50$, $J=1$). Note that the DMRG diagonalization (solid lines) gets closer to the Born-Oppenheimer prediction (dashed lines) for decreasing $\delta$, whereas the Silbey-Harris ansatz (dashed-dotted lines) improves for bigger values of the boson energy. The step for the derivatives in all cases is the same and stems from the precision used in the DMRG diagonalization: $\Delta g = 0.02 \cdot J$.}
    \label{Nder_comparisons}
	\end{flushleft}
\end{figure}

However, our finite-size results are consistent with the occurrence of a first order phase transition in the regime of slow boson dynamics. To assess this phenomenology, we focus on the behaviour of the mean boson number $n$. As depicted in Fig. \ref{Ncomparisons}, $n$ shows a sudden change when $\delta$ lies deep in the regime $\delta \ll J$, whereas the discontinuity vanishes for $\delta \geq J$ (we set units such that $J=1$). To quantify better that discontinuity and to place accurately the position of the phase transition, we have computed the numerical derivative of $n$ as a function of $g$, see Fig. \ref{Nder_comparisons}. We observe that for $\delta$ below $J$ the numerical derivative
inversely scales with $\delta$. This results is consistent with the sudden change in the ground state between the $n \simeq 0$, F phase, to the displaced vacuum of the DF phase, where $n=\alpha^2_0 \sim \delta^{-2}$ according to Eq. (\ref{number.of.bosons.bo}). Increasing the values of $\delta$ leads to a disappearance of any peak in the numerical derivative. We conclude then that the discontinuous behaviour is only unveiled in the limit $\delta \ll J$, because any signature is lost when $\delta \geq J$.

We have also compared the exact results with the variational approaches. Let us start by checking the accuracy of the Born-Oppenheimer approximation, which works in the limit $\delta \ll J, g$. To this end, we look for a closer resemblance between the BO solution and the exact diagonalization for decreasing values of $\delta$ (cf. Fig. \ref{Ncomparisons}). Accordingly, we see that the smaller the $\delta$, the nearer the BO prediction for the number of bosons $n$ lies to the DMRG observable. This is also true in the case of the derivative of the number of bosons where, in contrast to the Silbey-Harris ansatz, the BO approximation quantitatively predicts the height of the derivative when $\delta \to 0$.
\begin{figure}[h!]
	\begin{flushleft}
	\includegraphics[width=3.5in]{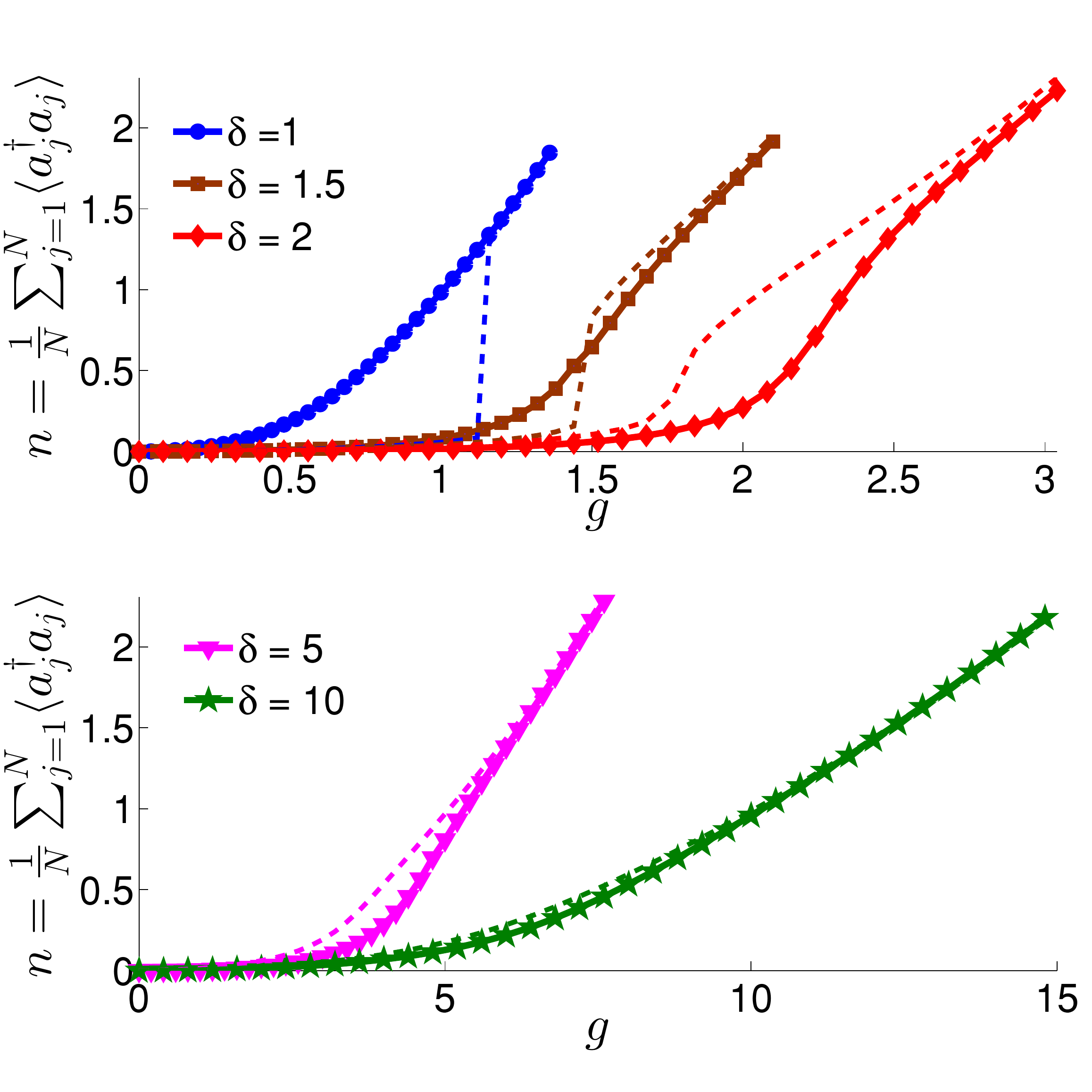}
    \caption{DMRG average boson number (solid lines with symbols) vs. Silbey-Harris ansatz (dashed line) for $\delta>J$ ($N=50$, $J=1$).}
    \label{N_big_delta}
	\end{flushleft}
\end{figure}

Regarding the Silbey-Harris approach, Fig. \ref{Ncomparisons} shows that it correctly describes the existence of the discontinuity. However, this solution must also give a suitable description of the phase with $\delta \gg J$, as we know that the dressed-ferromagnetic phase consists of a displaced state. We have therefore run simulations for bigger values of $\delta$ and $g$ (cf. Fig. \ref{N_big_delta}) and compared them with the SH ansatz, that effectively coincides with the exact solution when $\delta, g \gg J$.

\begin{figure}[h!]
	\begin{flushleft}
	\includegraphics[width=3.7in]{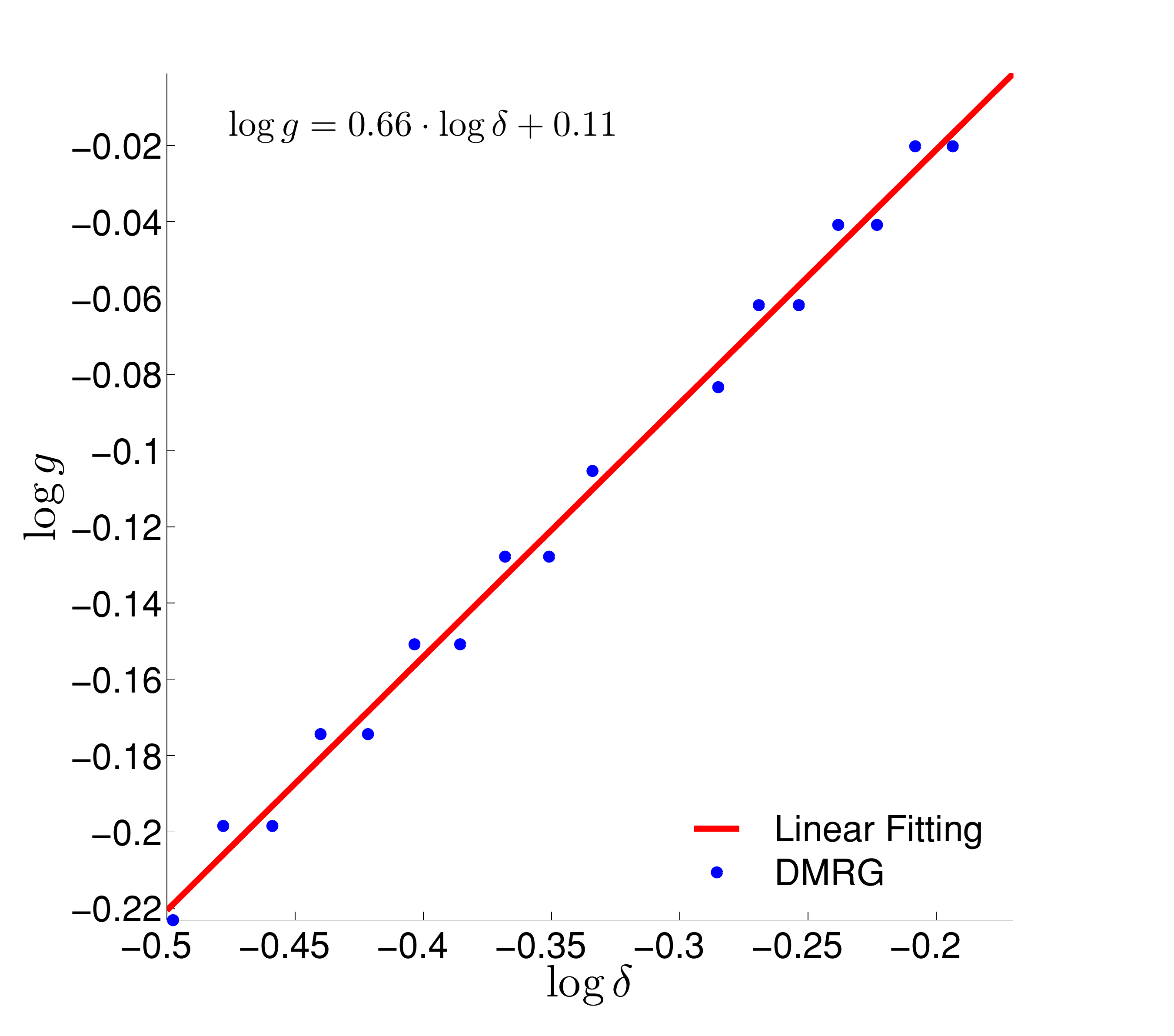}
	\caption{Linear fit of the critical line $g_{\rm c} (\delta)$ from the DMRG results ($N=50$, $J=1$). We depict both the logarithm of $\delta, g$. The result is coherent with a power law decay $g_{\rm c} \sim \delta^{\alpha}$, with $\alpha \simeq 0.66$.}
	\label{fitting}
	\end{flushleft}
\end{figure}

In Fig. \ref{fitting} we present the scaling of the critical line with the parameter $\delta$, in the regime $\delta < J$. It has been obtained from the position of the maxima of the derivatives of $n$ as a function of $g$, for different values of $\delta$. This allows us to calculate the function $g_{\rm c}(\delta)$, defining the critical line. Our results yield
a power law, e.g., $g_{\rm c} \sim \delta^{\alpha}$ for fixed $J$, with the exponent $\alpha = 0.66$. 
The quasi-exact numerical result departs from the BO approximation, that predicts $g_{\rm c}= \sqrt{\delta J}$, that is, $\alpha =1/2$. 

We have also studied signatures of the first order phase transition in the correlation length. Let us define the spin correlation functions,
\begin{equation}
C_z(i,j) = \langle \sigma_i^z \sigma_j^z \rangle-\langle \sigma_i^z \rangle \langle \sigma_j^z \rangle.
\end{equation}

\begin{figure}[h!]
	\begin{flushleft}
	\includegraphics[width=3.5in]{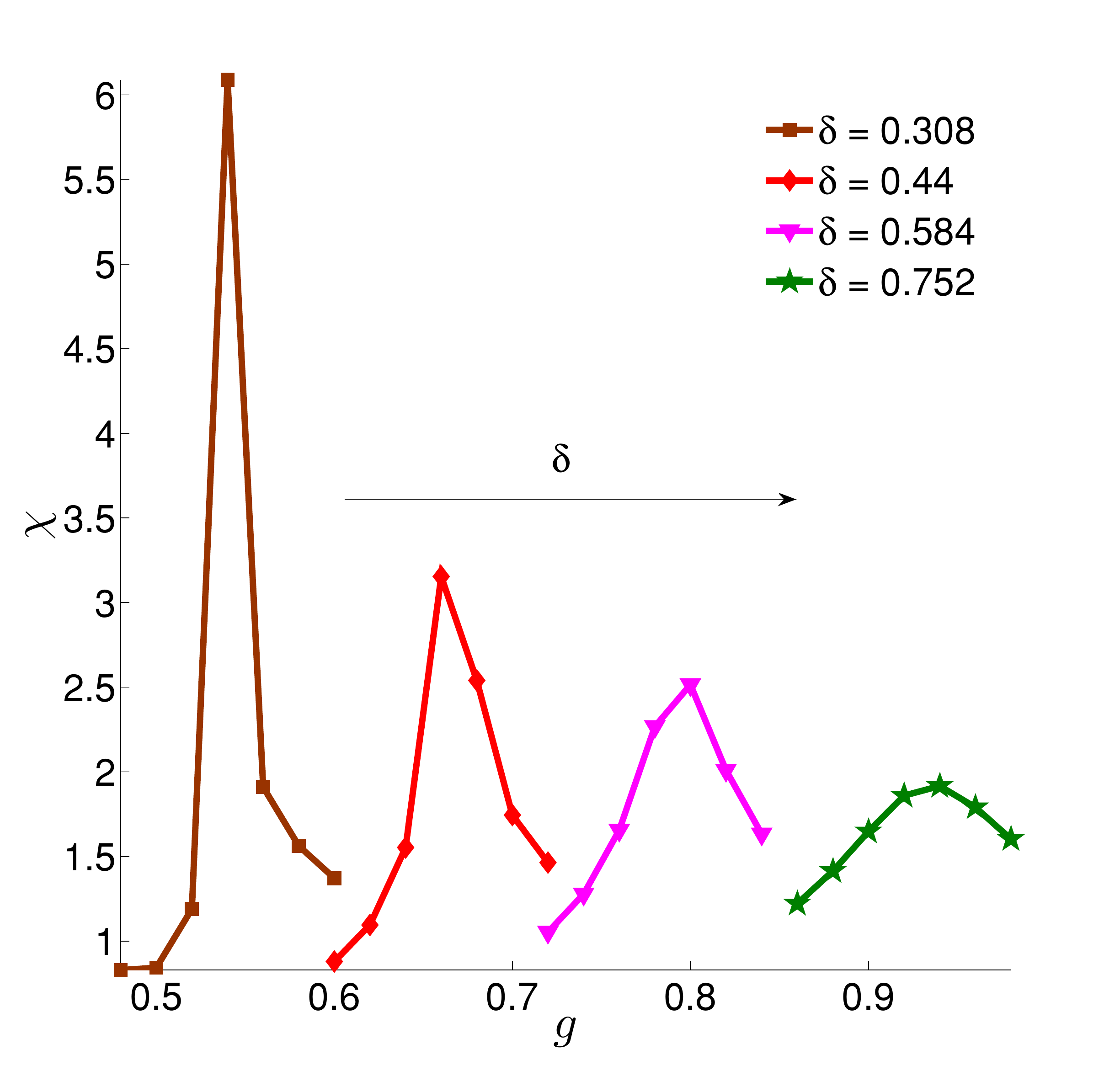}
	\caption{Correlation lengths $\chi$ obtained as the slope of $-\log (\sum_{j=1}^{25} C_z(25,25 + j)/N)$, across the critical line as a function of $g$, for different values of $\delta$ ($N=50$, $J=1$).}
	\label{corr.vs.g}
	\end{flushleft}
\end{figure}

\begin{figure}[h!]
	\begin{flushleft}
	\includegraphics[width=3.5in]{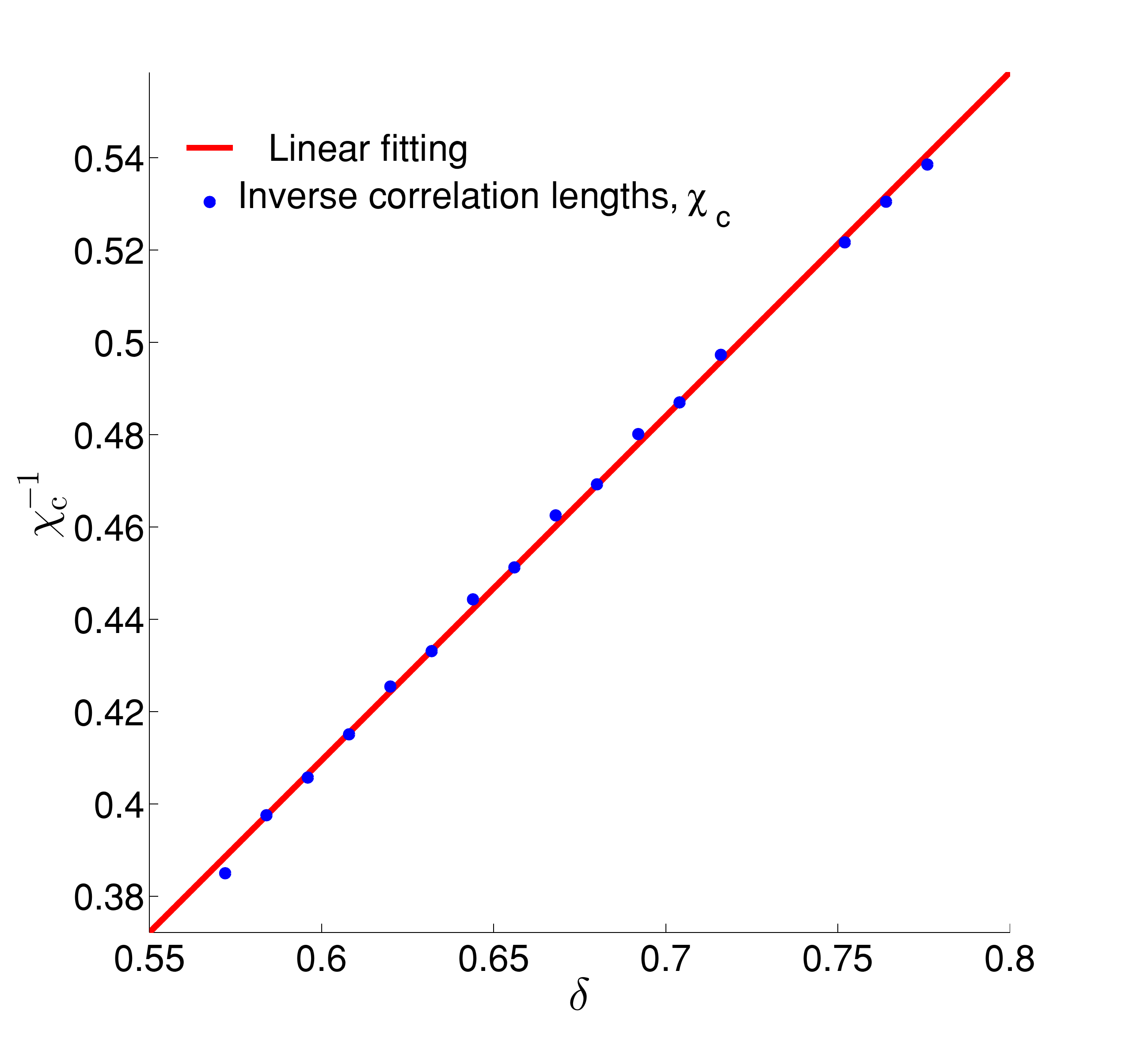}
	\caption{Fitting of $\chi^{-1}_{\rm c}$ to a line, from the DMRG with $N=50$, $J=1$. The results are coherent with the fact that the gap $\Delta$ scales linearly with the bosons energy $\delta$ along the critical line.}
	\label{correlations}
	\end{flushleft}
\end{figure}

Our calculations show that $C_z(i,j) \propto e^{- |i - j|/\chi}$ along the whole phase diagram, where $\chi$ is the correlation length. The exponential decay is observed even close to the first order phase transition in the regime $\delta < J$. This is consistent with our picture of the transition as a level crossing: F and DF states are both close to eigenstates of $H_{\rm IR}$ at the critical point, and both of them show exponentially decaying correlations. This is in clear contrast with what one would expect in a second order phase transition \cite{SachdevBook}. On the critical line, $\delta$ can be identified as the energy gap separating the ground state sector from the lowest energy excitations. We thus expect that the correlation length on the critical line, $\chi_{\rm c}$, must be a decreasing function of $\delta$. Our DMRG calculations confirm this picture (Fig. \ref{corr.vs.g}), and yield the scaling $\chi_{\rm c} \propto 1/\delta$ (Fig. \ref{correlations}). 

\section{Implementation of the Ising-Rabi Lattice Hamiltonian with trapped ions}

In this section we discuss an eventual realization of the IR Hamiltonian in state-of-the-art trapped ion set-ups, where highly accurate state preparation and readout is currently achievable \cite{Schneider12rpp}. In these systems, two electronic levels of every ion are chosen to be regarded as the spin degrees of freedom, whereas the quantized oscillations of the ions (phonons) give rise to the bosons. Then, spins and phonons are coupled through optical forces. We devise using a linear array of microtraps \cite{Home09sci,Schmied09prl,Kumph11njp,Hensinger14natcom} instead of the more usual Paul traps \cite{Leibfried03rmp}. Individual traps are specially suitable for our purpose, as their frequencies can be independently tuned, and the motion of every ion can be made resonant with a different laser force.

There exist as well other experimental set-ups for the simulation of the Hamiltonian $H_{\rm IR}$, such as superconducting qubits \cite{Houck12natphys} or Rydberg atoms \cite{Molmer10rmp}. For example, in this latter case the spin-spin interaction --between the two level systems made up of the ground and (very high) excited state of every atom in the sample-- is directly induced by the electromagnetic interaction between the electronic states. The spin-boson coupling can be introduced by the action of lasers as in the trapped ions experiments.

\subsection{Description of the set-up}

We discuss the set-up in three parts, consisting of the three different terms in $H_{\rm IR}$. The reader is referred to \cite{Schneider12rpp,Leibfried03rmp} for further details of the implementation.

\subsubsection{Phonon Hamiltonian}

Let us assume a linear array of traps forming an ion chain along the $z$ axis. Furthermore, we consider a constant separation $d_0$ between traps. The (quantum) position of the ions can be written as
\begin{equation}
\vec{r}_j = \delta r_{x,j} \hat{x} + \delta r_{y,j} \hat{y}+(z_j^0+\delta r_{z,j}) \hat{z},
\end{equation}
where operators $\delta r_{\alpha,j}$ stand for the displacements off their equilibrium coordinates $(0,0,z^0_{j})$. Motion along the $y$ axis is not relevant for the simulation, and will be omitted in the following. Displacements between different directions are decoupled assuming effectively harmonic trapping potentials, and approximating the Coulomb interaction up to second order in $\delta r_{\alpha,j}$. Therefore, ions are subjected to the effective potential
\begin{equation}
V = \frac{1}{2} m \sum_{\alpha,j} {\omega}_{\alpha, j}^2 \delta r_{\alpha,j}^2 - \sum_{\substack{\alpha \\j,l\neq j}} \frac{c_{\alpha}e^2/2}{|z_j^0-z_l^0|^3} (\delta r_{\alpha,j}-\delta r_{\alpha,l})^2.
\end{equation}

In this expression $c_{x}=1,c_{z}=-2$, $m$ is the ion mass, $e$ the electron charge in CGS units, and $\omega_{\alpha, j}$ are trap dependent frequencies. The corresponding Hamiltonian can be canonically quantized expressing the positions and momenta in terms of creation and annihilation operators, so that
\cite{Porras04prl,Deng08pra,Haze12pra},  (we take $\hbar = 1$)
\begin{equation}
H_{\rm phonon}=\sum_{\alpha} \{\sum_j {\omega}_{\alpha,j} a^{\dagger}_{\alpha,j}a_{\alpha,j} + \sum_{j}\sum_{l \neq j} t^{\alpha}_{j,l} a^{\dagger}_{\alpha,j}a_{\alpha,l}\},
\label{phonon.hamiltonian}
\end{equation}
where
\begin{equation}
t^{\alpha}_{j,l} = \sum_{l\neq j}\frac{c_{\alpha}e^2}{2m(\omega_{\alpha,j}\omega_{\alpha,l})^{1/2}|z_j^0-z_l^0|^3}.
\label{hopping.strengths}
\end{equation}

In (\ref{phonon.hamiltonian}) is already assumed that $\omega_{\alpha,j}\gg t^{\alpha}_{j,l}$, such that corrections to on-site frequencies stemming from the dipolar interaction, or phonon non-conserving terms, are negligible.

We need to get rid of the hopping terms $a^{\dagger}_{\alpha,j}a_{\alpha,l}$ in $H_{\rm phonon}$, at least for one direction $\alpha$, in order to give rise to the local boson contribution in the IR Hamiltonian, e.g., $\delta \sum_j a^{\dagger}_j a_j$. Let us choose for this purpose the transversal modes along the $x$ axis. Then, we propose using different trap frequencies $\omega_{x,j}$ to make hopping events in (\ref{phonon.hamiltonian}) fast rotating compared to the on-site energies. Specifically, if ions $j$ and $l$ are subjected to frequencies $\omega_{x,j}$ and $\omega_{x,l}$, the terms $a^{\dagger}_{\alpha,j}a_{\alpha,l}$ would rotate with $ \exp[-it(\omega_{x,j}-\omega_{x,l})]$ in the interaction picture for the motion. The so-called Rotating Wave Approximation (RWA) prescribes that such terms are negligible as long as $t^x_{j,l} \ll |\omega_{x,j}-\omega_{x,l}|$. Assuming this is the case, hopping terms in $H_{\rm phonon}$ can be safely ignored, and we are led to the term
\begin{equation}
H_x = \sum_{j=1}^N \omega_{x,j} a^{\dagger}_{x,j} a_{x,j}
\end{equation}
we were aiming for. The common frequency $\omega_{x,j}\to \delta$ can be achieved by means of local laser detunings, discussed later on. The motional coupling between different traps decays fast as a function of the ion-ion distance, $t_{j,l}^{\alpha} \sim 1/|z^0_j-z^0_l|^3$. Thus, it is only necessary to eliminate the coupling between nearest or next-to-nearest neighbour ions, since longer-range terms will give negligible contributions.


Regarding the motion in the $z$ direction, we set $\omega_{z,j}\to \omega_z$. Since trap frequencies along $x$ and $z$ are independently and locally tunable, this choice can be made at no expense of the previous discussion. The Hamiltonian (\ref{phonon.hamiltonian}) reads then 
\begin{equation}
H_{z}=\sum_{n=0}^{N-1} \omega_{z,n} a^{\dagger}_{z,n} a_{z,n}
\end{equation} 
in the basis of collective modes of motion $a_{z,n} = \sum_{j=1}^N M^z_{j,n} a_{z,j}$, with normal frequencies $\omega_{z,n}$. This term does not occur in $H_{\rm IR}$ as we aim at a regime where $\langle a^{\dagger}_{z,n} a_{z,n}\rangle \simeq 0$. Nonetheless, their (virtual) exchange creates the spin-spin interaction \cite{Porras04prl}.
\begin{figure}[h!]
	\includegraphics[width=3.3in]{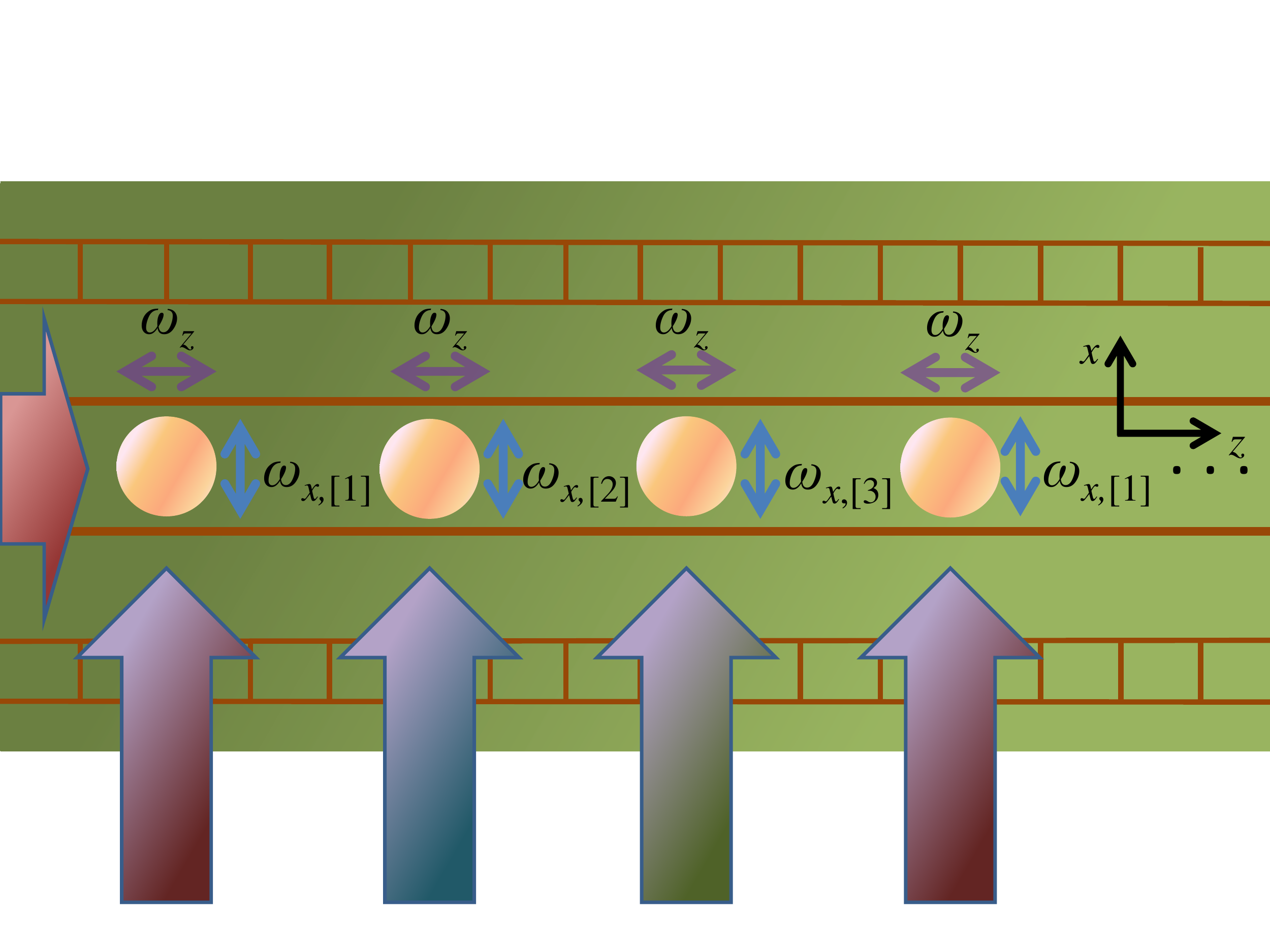}
    \caption{Scheme representing trapped ions in linear array of microtraps as electrodes printed over a surface. Solid arrows represent the laser fields acting on the ion chain. We indicate the trap frequencies at every site.}
    \label{implementation}
\end{figure}

\subsubsection{Spin-spin interaction}

Implementing the exchange term of $H_{\rm IR}$ relies on inducing a spin-spin effective coupling. Let us assume a laser field, lying along the direction of the linear array of traps, with momentum $\Delta k_{z}$ and frequency $\omega^{\rm L}_z = \omega_{z,n}-\delta_{z,n}$. Here $\delta_{z,n}$ stands for the laser detuning from the $n$ axial normal mode. Differential a.c. Stark shifts stemming from the off-resonant components of the atom-light interaction give rise to a spin-dependent $\sigma^z$-force \cite{Wineland03ptrs} of the form
\begin{equation}
H_{\rm z-force} = g_z \sum_{j,n} \sigma_j^z \left(M_{j,n}^z a_{z,n} + {\rm H.c.} \right),
\label{z.force}
\end{equation}
with coupling strength $g_{z}$. This Hamiltonian is time independent because we have moved to a rotating frame, where phonon frequencies are shifted, $\omega_{z,n}\to \omega_{z,n}-\omega^{\rm L}_z = \delta_{z,n}$. Performing a transformation to a displaced basis (see e.g. \cite{Porras04prl}), the previous force takes the form of the effective spin-spin interaction 
\begin{equation}
H_{\rm exchange} = \sum_{j,l} J_{j,l} \sigma^z_j \sigma^z_l,\, J_{j,l} \simeq -\frac{J}{|j-l|^3},
\label{exchange.hamiltonian}
\end{equation}
for suitable detunings and shapes of the axial modes spectrum. This interaction acts effectively as a first neighbours ferromagnetic coupling of magnitude $J$.

\subsubsection{Local spin-phonon coupling}

Local spin-phonon couplings in $H_{\rm IR}$ require driving simultaneously red and blue sideband transitions \cite{Leibfried03rmp} for the transversal oscillations. However, as we have already discussed, $\omega_{x,j}$ are different among close traps. This means that matching the resonance conditions for the spin-dependent forces requires as many laser wavelengths as different trapping frequencies. Let us consider the array of traps as consisting of $N/n,\, n\in \mathbb{N}$ sequential sets of traps. Within these, neighbouring traps frequencies are different. We set a constant difference between one trap and the next, $\omega_{x,j}-\omega_{x,j+1}=\Delta \omega_x$. All the sets have the same arrangement of $n$ frequencies, and they appear one after the other along the chain. 
Let us call these frequencies $\omega_{x,1},\ldots, \omega_{x,n}$.
Any frequency can be written then as $\omega_{x,[j]}$, where $[j] = (j-1) \bmod n + 1$. 
Now, we apply $n$ laser fields transversally to the chain, with mutual detunings $\Delta \omega_x, \ldots, (n-1)\Delta \omega_x$. Because of this frequency difference, they can address the whole chain at the same time. In this way, the matching condition only happens between a given laser with, let us say $\omega_{x,[j]}^{\rm L} =\omega_0 + \omega_{x,[j]} - \delta_{x,[j]}$, and the ions that are trapped at frequencies $\omega_{x,[j]}$ ($\omega_0$ is the spin transition frequency). This gives rise to the $\sigma^x$-force
\begin{equation}
H_{\rm x-force}(t)= g \sum_{j=1}^N \sigma^x_j(a_{x,j}^{\dagger} e^{i\delta_{x,[j]} t}+a_{j,x}e^{-i\delta_{x,[j]} t}),
\end{equation}
where $g = i \Omega_{x,[j]} \eta_{x,[j]}$, the laser Rabi frequency and Lamb-Dicke parameters of the coupling, respectively. We rely on the local dependence of $\Omega_{x,[j]}$ to achieve a homogeneous $g$ along the chain, as $\eta_{x,[j]}$ depend on the on-site trap frequencies. 

Finally, moving into a rotating frame with frequencies $\omega_{x,[j]}^{\rm L}$, we get ${\omega}_{x,[j]}\to \delta_{x,[j]}$ in $H_x$, and $H_{\rm x-force} (t)\to H_{\rm x-force}(0)$, so that
\begin{equation}
H_{\rm x-force} = g \sum_{j=1}^N \sigma^x_j(a_{x,j}^{\dagger} +a_{j,x}).
\end{equation}

Since laser detunings are site-dependent, they can be shifted to give common on-site phonon energies $\delta,\, \forall j$, which leads to
\begin{equation}
H_x = \sum_{j=1}^N \delta a_{x,j}^{\dagger} a_{x,j}
\end{equation}
as the effective phonon energy contribution.

The IR Hamiltonian is eventually implemented as the sum of $H_x,\,H_{\rm exchange}$ and $H_{\rm x-force}$.

\subsection{Trapped ions experimental parameters}

We consider the traps separated by a distance $d_0=30 \, \mu$m, every of them containing one $\mathrm{^9 Be^+}$ ion. We estimate $|z^0_j-z^0_l|=d_0$ in (\ref{hopping.strengths}). Eq. (\ref{z.force}) holds only if $\max\eta_{z,n} = \Delta {k}_{z}/{\sqrt{2m {\omega}_{z,n=0}}}\ll 1$. We propose a common $\omega_z = 500\,(2\pi)$ KHz  for all traps, which leads to $t^z_{j,j+1}\simeq 29 \,(2\pi)$ KHz and to $\omega_{z,n=0}\simeq 431\,(2\pi)$ KHz for the ground state COM frequency of the axial modes band. Therefore, a laser wavelength $\lambda^z_{\rm L} \simeq 870$ nm would give $ \eta_{z,n=0} \simeq 0.26$ for beams on axis with the traps.

The magnitude of the exchange in (\ref{exchange.hamiltonian}) is $J \simeq t^z_{j,j+1} g_z^2/\delta_{z,n=0}^2$, where $g_z$ has typical values $100\,(2\pi)$ KHz \cite{Schneider12rpp}, whereas we impose $\delta_{z,n=0}\simeq 2g_z$ in order to neglect residual spin-phonon couplings \cite{Porras04prl}. This renders the value $J\simeq 7\,(2\pi)$ KHz, which is the lowest energy scale involved in the simulation.

The number of different frequencies $\omega_{x,j}$ fixes an error bound for the simulation. Ions trapped at equal frequencies are coupled by a residual dipolar interaction, whose magnitude is $\max_j (t^x_{j,j+n})$. We aim at making it very small with respect to the rest of parameters in $H_{\rm IR}$. Then, processes with energies $\sim \max_j (t^x_{j,j+n})$ will be systematically neglected.
For the sake of concreteness, we address the example of $n=3$. Assuming $\omega_{x,[1]}=10\,(2\pi)$ MHz, $\omega_{x,[2]}=9\,(2\pi)$ MHz, and $\omega_{x,[3]}=8\,(2\pi)$ MHz, we have  $\max_j(t^x_{j,j+1})\simeq 0.9\, (2\pi)$ KHz. This amount scales with the distance, so that $\max_j(t^x_{j,j+n}) \sim \max_j(t^x_{j,j+1})/n^3 \simeq 33\,(2\pi)$ Hz. Accordingly, we prescribe $\delta, g, J \gg \max_j(t^x_{j,j+n})$ as the condition to be fulfilled to safely neglect residual couplings. Furthermore, with the former choice of parameters, the RWA condition is also fulfilled, as $\max_{j,l}(t^x_{j,l}/|\omega_{x,j}-\omega_{x,l}|) \simeq 10^{-3}, l = j+1,\cdots,n$.

Regarding the spin-boson interaction, we consider laser beams with effective wavelength $\lambda^x_{\rm L} \simeq 320$ nm acting transversely to the traps' axis.
Thus, the Lamb-Dicke parameters are $\max\eta_{x,j}\simeq 0.16$. Typical values for $g$ are again of the order of $100\,(2\pi)$ KHz. The energy of the transverse phonons is set by locally detuning from $\bar{\omega}_{x,j}$ to the common value $\delta$ for every site, and it can chosen such that $\delta\sim g$, as we have theoretically studied.

In order to probe the phase transition we propose preparing the ferromagnetic phase by cooling to the ground state of the phonons, while optical pumping to the $\bigotimes_j \left|\downarrow_z\right\rangle_j$ spin state, where $\left|\downarrow_z\right\rangle_j$ is one of the qubit states. An adiabatic protocol crossing the critical line would require evolution times of the order of the inverse of the smallest of the parameters, which lies around $t^{-1}\sim 23\,\mu$s. 

\section{Conclusions}

We have introduced the Ising Rabi Lattice model, that consists of a generalization of the single particle Rabi model that includes Ising couplings between spins. Our model departs from the Ising universality class, and presents a discrete gauge symmetry. We have used several approximations and perturbative arguments that predict a quantum phase diagram divided in two parts: (i) Slow boson regime ($\delta \ll J$) in which a first order phase transition separates a ferromagnetic phase from a phase with a dressed-ferromagnetic phase. (ii) Fast boson regime ($\delta \gg J$), were the transition between the F and DF phases is continuous. This picture is consistent with quasi-exact numerical calculations with the DMRG method. Our model can be implemented with trapped ions in arrays of microtraps, leading to the implementation of gauge symmetries and first order phase transitions in this system.

\section*{Acknowledgements}

P.N. acknowledges H. Takahashi for its help in details of the implementation. D.P. acknowleges J.J. Garc\'{\i}a-Ripoll for suggesting the application of the Silbey-Harris variational approach in this work. Work supported by the EU Marie Curie Career Integration Grant 630955 NewFQS.

\appendix

\section{Survival of the two-fold degeneracy up to any finite order in perturbation theory}
\label{App:A}

In the $g = 0$ limit of $H_{\rm IR}$, its ground states are those of
\begin{equation}
H^0_{\rm F} = \delta \sum_{j=1}^{N} a_j^{\dagger} a_j - J \sum_{j=1}^{N-1}  \sigma_j^z \sigma_{j+1}^z,
\end{equation}
which fulfil
\begin{equation}
|\phi_{\rm F} \rangle \in \operatorname{l.c.}
\left\{ |0\rangle_{\rm b} \bigotimes_{j=1}^N \left|\uparrow_{z} \right\rangle, |0\rangle_{\rm b} \bigotimes_{j=1}^N \left|\downarrow_{z} \right\rangle \right\},
\label{ferromagnet}
\end{equation}
where $\operatorname{l.c.}$ stands for every independent linear combination of these vectors. Because of the two-fold degeneracy, we are allowed to choose as ground states any two elements of (\ref{ferromagnet}), so for the sake of simplicity we will consider that $|\phi_{\rm F} \rangle$ is any of the ferromagnetic orders
\begin{equation}
|\phi_{\uparrow} \rangle := |0\rangle_{\rm b} \bigotimes_{j=1}^N \left|\uparrow_{z} \right\rangle,  |\phi_{\downarrow} \rangle := |0\rangle_{\rm b} \bigotimes_{j=1}^N \left|\downarrow_{z} \right\rangle.
\label{ground_states}
\end{equation}

If we consider now the perturbation
\begin{equation}
H'_{\rm F} = g \sum_{j=1}^{N} \sigma_j^x (a_j^{\dagger} + a_j),
\end{equation}
the first step in building corrections to $|\phi_{\rm F} \rangle$ consists in finding the correct linear combination, i.e., the weights $c_{\uparrow}, c_{\downarrow}$ in $|\phi_{\rm F} \rangle\,(g \to 0)\,\rangle  = c_{\uparrow} |\phi_{\uparrow} \rangle  + c_{\downarrow} |\phi_{\downarrow} \rangle$, that continuously matches the ground state for $g = 0$.  This is accomplished by means of degenerate perturbation theory (see \cite{SchiffBook} for an account of the $Brillouin-Wigner$ approach), which studies the effect of the perturbation within the degenerate subspace (\ref{ground_states}): if $H'_{\rm F}$ is such that it lifts the degeneracy at a given order $n$, this procedure provides two new eigenvectors whose energies are the eigenvalues $E_{\rm GS}^{(n)}$ of the following secular equations
\begin{widetext}
\begin{equation}
\left\{ \begin{array}{l}
 E_{\rm GS}^{(n)} c_{\uparrow} = E_{\uparrow} c_{\uparrow} + \langle \phi_{\uparrow} | \displaystyle \frac{H'_{\rm F}}{1 - R_{\rm GS} H'_{\rm F}} |\phi_{\uparrow} \rangle c_{\uparrow} +  \langle \phi_{\uparrow} | \displaystyle \frac{H'_{\rm F}}{1 - R_{\rm GS} H'_{\rm F}} |\phi_{\downarrow} \rangle c_{\downarrow},\\[0.5cm]
E_{\rm GS}^{(n)} c_{\downarrow} = E_{\downarrow} c_{\downarrow} + \langle \phi_{\downarrow} | \displaystyle \frac{H'_{\rm F}}{1 - R_{\rm GS} H'_{\rm F}} |\phi_{\downarrow} \rangle c_{\downarrow} + \langle \phi_{\downarrow} | \displaystyle \frac{H'_{\rm F}}{1 - R_{\rm GS} H'_{\rm F}} |\phi_{\uparrow} \rangle c_{\uparrow} .
\end{array} \right.
\label{N_perturbation}
\end{equation}
\end{widetext}

In the former expressions $E_{\uparrow} = E_{\downarrow}$ are the ground state energies for $g=0$ and $R_{\rm GS}=(E_{\rm GS}^{(n)}-H^0_{\rm F})^{-1} \cdot (1 - |\phi_{\rm F}\rangle \langle \phi_{\rm F}|)$ is known as the {\it resolvent}.

In the present case, we are going to show that the degeneracy is not lifted at any finite order in perturbation theory. We note that equations (\ref{N_perturbation}) give only one solution for $E_{\rm GS}^{(n)}$ if and only if both
\begin{equation}
\left\{ \begin{array}{ll} 
\langle \phi_{\uparrow} |\displaystyle  \frac{H'_{\rm F}}{1 - R_{\rm GS} H'_{\rm F}} |\phi_{\uparrow}\rangle = \langle \phi_{\downarrow} | \displaystyle \frac{H'_{\rm F}}{1 - R_{\rm GS} H'_{\rm F}} |\phi_{\downarrow} \rangle, \\[0.5cm]
\langle \phi_{\downarrow} | \displaystyle \frac{H'_{\rm F}}{1 - R_{\rm GS} H'_{\rm F}} |\phi_{\uparrow} \rangle = 0,
\end{array} \right.
\label{conditions}
\end{equation}
{\it do hold}. The first of these conditions is trivially fulfilled because of parity arguments, but the second must be computed explicitly. It turns out that it holds as well, because all the matrix elements in
\begin{equation}
\langle \phi_{\downarrow} | \frac{H'_{\rm F}}{1 - R_{\rm GS} H'_{\rm F}} |\phi_{\uparrow} \rangle \overset{O(n)}{=} \sum_{k=0}^n \langle \phi_{\downarrow} | H'_{\rm F} \left(R_{\rm GS} H'_{\rm F}\right)^k |\phi_{\uparrow}\rangle
\end{equation}
are zero. To show this, let us write the generic form of a n-th order contribution to the previous sum (denominators can be neglected for this discussion),
\begin{equation}
g^n  \bigotimes_{j=1}^N \,_{\rm b} \langle 0 | (a_j^{\dagger} + a_j)^{n_j} |0\rangle_{\rm b} \left\langle \downarrow_z \right| (\sigma_j^x)^{n_j} \left| \uparrow_{z}\right\rangle,
\end{equation}
with $\sum_{j=1}^N n_j = n$. We note that the boson displacement terms are diagonal only in the event of even values of every $n_j$. However, only an odd value of all the $n_j$ would give a non-zero contribution from the spin part, because in another case the tunneling matrices are equal to the unit matrix. Since both contributions cannot be simultaneously different from zero, we conclude that the second condition in (\ref{conditions}) is also fulfilled.

According to these previous considerations, the two-fold degeneracy in the ground state is not lifted in any finite order of perturbation theory, which in turns translates into the fact that degeneracy remains for any finite value of $g$ \cite{SachdevBook}. Therefore, perturbative corrections must be carried upon any element of (\ref{ground_states}) by means of conventional non-degenerate perturbation theory.

\bibliography{NevadoPorrasPRA_biblio}

\begin{thebibliography}{31}%
\makeatletter
\providecommand \@ifxundefined [1]{%
 \@ifx{#1\undefined}
}%
\providecommand \@ifnum [1]{%
 \ifnum #1\expandafter \@firstoftwo
 \else \expandafter \@secondoftwo
 \fi
}%
\providecommand \@ifx [1]{%
 \ifx #1\expandafter \@firstoftwo
 \else \expandafter \@secondoftwo
 \fi
}%
\providecommand \natexlab [1]{#1}%
\providecommand \enquote  [1]{``#1''}%
\providecommand \bibnamefont  [1]{#1}%
\providecommand \bibfnamefont [1]{#1}%
\providecommand \citenamefont [1]{#1}%
\providecommand \href@noop [0]{\@secondoftwo}%
\providecommand \href [0]{\begingroup \@sanitize@url \@href}%
\providecommand \@href[1]{\@@startlink{#1}\@@href}%
\providecommand \@@href[1]{\endgroup#1\@@endlink}%
\providecommand \@sanitize@url [0]{\catcode `\\12\catcode `\$12\catcode
  `\&12\catcode `\#12\catcode `\^12\catcode `\_12\catcode `\%12\relax}%
\providecommand \@@startlink[1]{}%
\providecommand \@@endlink[0]{}%
\providecommand \url  [0]{\begingroup\@sanitize@url \@url }%
\providecommand \@url [1]{\endgroup\@href {#1}{\urlprefix }}%
\providecommand \urlprefix  [0]{URL }%
\providecommand \Eprint [0]{\href }%
\providecommand \doibase [0]{http://dx.doi.org/}%
\providecommand \selectlanguage [0]{\@gobble}%
\providecommand \bibinfo  [0]{\@secondoftwo}%
\providecommand \bibfield  [0]{\@secondoftwo}%
\providecommand \translation [1]{[#1]}%
\providecommand \BibitemOpen [0]{}%
\providecommand \bibitemStop [0]{}%
\providecommand \bibitemNoStop [0]{.\EOS\space}%
\providecommand \EOS [0]{\spacefactor3000\relax}%
\providecommand \BibitemShut  [1]{\csname bibitem#1\endcsname}%
\let\auto@bib@innerbib\@empty
\bibitem [{\citenamefont {{Cirac}}\ and\ \citenamefont
  {{Zoller}}(2012)}]{Cirac12natphys}%
  \BibitemOpen
  \bibfield  {author} {\bibinfo {author} {\bibfnamefont {J.~I.}\ \bibnamefont
  {{Cirac}}}\ and\ \bibinfo {author} {\bibfnamefont {P.}~\bibnamefont
  {{Zoller}}},\ }\href {\doibase 10.1038/nphys2275} {\bibfield  {journal}
  {\bibinfo  {journal} {Nature Physics}\ }\textbf {\bibinfo {volume} {8}},\
  \bibinfo {pages} {264} (\bibinfo {year} {2012})}\BibitemShut {NoStop}%
\bibitem [{\citenamefont {{Friedenauer}}\ \emph {et~al.}(2008)\citenamefont
  {{Friedenauer}}, \citenamefont {{Schmitz}}, \citenamefont {{Glueckert}},
  \citenamefont {{Porras}},\ and\ \citenamefont
  {{Schaetz}}}]{Friedenauer08natphys}%
  \BibitemOpen
  \bibfield  {author} {\bibinfo {author} {\bibfnamefont {A.}~\bibnamefont
  {{Friedenauer}}}, \bibinfo {author} {\bibfnamefont {H.}~\bibnamefont
  {{Schmitz}}}, \bibinfo {author} {\bibfnamefont {J.~T.}\ \bibnamefont
  {{Glueckert}}}, \bibinfo {author} {\bibfnamefont {D.}~\bibnamefont
  {{Porras}}}, \ and\ \bibinfo {author} {\bibfnamefont {T.}~\bibnamefont
  {{Schaetz}}},\ }\href {\doibase 10.1038/nphys1032} {\bibfield  {journal}
  {\bibinfo  {journal} {Nature Physics}\ }\textbf {\bibinfo {volume} {4}},\
  \bibinfo {pages} {757} (\bibinfo {year} {2008})}\BibitemShut {NoStop}%
\bibitem [{\citenamefont {Schneider}\ \emph {et~al.}(2012)\citenamefont
  {Schneider}, \citenamefont {Porras},\ and\ \citenamefont
  {Schaetz}}]{Schneider12rpp}%
  \BibitemOpen
  \bibfield  {author} {\bibinfo {author} {\bibfnamefont {C.}~\bibnamefont
  {Schneider}}, \bibinfo {author} {\bibfnamefont {D.}~\bibnamefont {Porras}}, \
  and\ \bibinfo {author} {\bibfnamefont {T.}~\bibnamefont {Schaetz}},\ }\href
  {http://stacks.iop.org/0034-4885/75/i=2/a=024401} {\bibfield  {journal}
  {\bibinfo  {journal} {Reports on Progress in Physics}\ }\textbf {\bibinfo
  {volume} {75}},\ \bibinfo {pages} {024401} (\bibinfo {year}
  {2012})}\BibitemShut {NoStop}%
\bibitem [{\citenamefont {{Blatt}}\ and\ \citenamefont
  {{Roos}}(2012)}]{Blatt12natphys}%
  \BibitemOpen
  \bibfield  {author} {\bibinfo {author} {\bibfnamefont {R.}~\bibnamefont
  {{Blatt}}}\ and\ \bibinfo {author} {\bibfnamefont {C.~F.}\ \bibnamefont
  {{Roos}}},\ }\href {\doibase 10.1038/nphys2252} {\bibfield  {journal}
  {\bibinfo  {journal} {Nature Physics}\ }\textbf {\bibinfo {volume} {8}},\
  \bibinfo {pages} {277} (\bibinfo {year} {2012})}\BibitemShut {NoStop}%
\bibitem [{\citenamefont {{Houck}}\ \emph {et~al.}(2012)\citenamefont
  {{Houck}}, \citenamefont {{T{\"u}reci}},\ and\ \citenamefont
  {{Koch}}}]{Houck12natphys}%
  \BibitemOpen
  \bibfield  {author} {\bibinfo {author} {\bibfnamefont {A.~A.}\ \bibnamefont
  {{Houck}}}, \bibinfo {author} {\bibfnamefont {H.~E.}\ \bibnamefont
  {{T{\"u}reci}}}, \ and\ \bibinfo {author} {\bibfnamefont {J.}~\bibnamefont
  {{Koch}}},\ }\href {\doibase 10.1038/nphys2251} {\bibfield  {journal}
  {\bibinfo  {journal} {Nature Physics}\ }\textbf {\bibinfo {volume} {8}},\
  \bibinfo {pages} {292} (\bibinfo {year} {2012})}\BibitemShut {NoStop}%
\bibitem [{\citenamefont {Porras}\ \emph {et~al.}(2012)\citenamefont {Porras},
  \citenamefont {Ivanov},\ and\ \citenamefont {Schmidt-Kaler}}]{Porras12prl}%
  \BibitemOpen
  \bibfield  {author} {\bibinfo {author} {\bibfnamefont {D.}~\bibnamefont
  {Porras}}, \bibinfo {author} {\bibfnamefont {P.}~\bibnamefont {Ivanov}}, \
  and\ \bibinfo {author} {\bibfnamefont {F.}~\bibnamefont {Schmidt-Kaler}},\
  }\href {\doibase 10.1103/PhysRevLett.108.235701} {\bibfield  {journal}
  {\bibinfo  {journal} {Phys. Rev. Lett.}\ }\textbf {\bibinfo {volume} {108}},\
  \bibinfo {pages} {235701} (\bibinfo {year} {2012})}\BibitemShut {NoStop}%
\bibitem [{\citenamefont {Schir\'o}\ \emph {et~al.}(2012)\citenamefont
  {Schir\'o}, \citenamefont {Bordyuh}, \citenamefont {\"Oztop},\ and\
  \citenamefont {T\"ureci}}]{Tureci12prl}%
  \BibitemOpen
  \bibfield  {author} {\bibinfo {author} {\bibfnamefont {M.}~\bibnamefont
  {Schir\'o}}, \bibinfo {author} {\bibfnamefont {M.}~\bibnamefont {Bordyuh}},
  \bibinfo {author} {\bibfnamefont {B.}~\bibnamefont {\"Oztop}}, \ and\
  \bibinfo {author} {\bibfnamefont {H.~E.}\ \bibnamefont {T\"ureci}},\ }\href
  {\doibase 10.1103/PhysRevLett.109.053601} {\bibfield  {journal} {\bibinfo
  {journal} {Phys. Rev. Lett.}\ }\textbf {\bibinfo {volume} {109}},\ \bibinfo
  {pages} {053601} (\bibinfo {year} {2012})}\BibitemShut {NoStop}%
\bibitem [{\citenamefont {Schmidt}\ and\ \citenamefont
  {Koch}(2013)}]{Koch13adp}%
  \BibitemOpen
  \bibfield  {author} {\bibinfo {author} {\bibfnamefont {S.}~\bibnamefont
  {Schmidt}}\ and\ \bibinfo {author} {\bibfnamefont {J.}~\bibnamefont {Koch}},\
  }\href {\doibase 10.1002/andp.201200261} {\bibfield  {journal} {\bibinfo
  {journal} {Annalen der Physik}\ }\textbf {\bibinfo {volume} {525}},\ \bibinfo
  {pages} {395} (\bibinfo {year} {2013})}\BibitemShut {NoStop}%
\bibitem [{\citenamefont {Kurcz}\ \emph {et~al.}(2014)\citenamefont {Kurcz},
  \citenamefont {Bermudez},\ and\ \citenamefont {Garcia-Ripoll}}]{Kurcz14prl}%
  \BibitemOpen
  \bibfield  {author} {\bibinfo {author} {\bibfnamefont {A.}~\bibnamefont
  {Kurcz}}, \bibinfo {author} {\bibfnamefont {A.}~\bibnamefont {Bermudez}}, \
  and\ \bibinfo {author} {\bibfnamefont {J.~J.}\ \bibnamefont
  {Garcia-Ripoll}},\ }\href {\doibase 10.1103/PhysRevLett.112.180405}
  {\bibfield  {journal} {\bibinfo  {journal} {Phys. Rev. Lett.}\ }\textbf
  {\bibinfo {volume} {112}},\ \bibinfo {pages} {180405} (\bibinfo {year}
  {2014})}\BibitemShut {NoStop}%
\bibitem [{\citenamefont {Home}\ \emph {et~al.}(2009)\citenamefont {Home},
  \citenamefont {Hanneke}, \citenamefont {Jost}, \citenamefont {Amini},
  \citenamefont {Leibfried},\ and\ \citenamefont {Wineland}}]{Home09sci}%
  \BibitemOpen
  \bibfield  {author} {\bibinfo {author} {\bibfnamefont {J.~P.}\ \bibnamefont
  {Home}}, \bibinfo {author} {\bibfnamefont {D.}~\bibnamefont {Hanneke}},
  \bibinfo {author} {\bibfnamefont {J.~D.}\ \bibnamefont {Jost}}, \bibinfo
  {author} {\bibfnamefont {J.~M.}\ \bibnamefont {Amini}}, \bibinfo {author}
  {\bibfnamefont {D.}~\bibnamefont {Leibfried}}, \ and\ \bibinfo {author}
  {\bibfnamefont {D.~J.}\ \bibnamefont {Wineland}},\ }\href {\doibase
  10.1126/science.1177077} {\bibfield  {journal} {\bibinfo  {journal}
  {Science}\ }\textbf {\bibinfo {volume} {325}},\ \bibinfo {pages} {1227}
  (\bibinfo {year} {2009})}\BibitemShut {NoStop}%
\bibitem [{\citenamefont {Schmied}\ \emph {et~al.}(2009)\citenamefont
  {Schmied}, \citenamefont {Wesenberg},\ and\ \citenamefont
  {Leibfried}}]{Schmied09prl}%
  \BibitemOpen
  \bibfield  {author} {\bibinfo {author} {\bibfnamefont {R.}~\bibnamefont
  {Schmied}}, \bibinfo {author} {\bibfnamefont {J.~H.}\ \bibnamefont
  {Wesenberg}}, \ and\ \bibinfo {author} {\bibfnamefont {D.}~\bibnamefont
  {Leibfried}},\ }\href {\doibase 10.1103/PhysRevLett.102.233002} {\bibfield
  {journal} {\bibinfo  {journal} {Phys. Rev. Lett.}\ }\textbf {\bibinfo
  {volume} {102}},\ \bibinfo {pages} {233002} (\bibinfo {year}
  {2009})}\BibitemShut {NoStop}%
\bibitem [{\citenamefont {{Kumph}}\ \emph {et~al.}(2011)\citenamefont
  {{Kumph}}, \citenamefont {{Brownnutt}},\ and\ \citenamefont
  {{Blatt}}}]{Kumph11njp}%
  \BibitemOpen
  \bibfield  {author} {\bibinfo {author} {\bibfnamefont {M.}~\bibnamefont
  {{Kumph}}}, \bibinfo {author} {\bibfnamefont {M.}~\bibnamefont
  {{Brownnutt}}}, \ and\ \bibinfo {author} {\bibfnamefont {R.}~\bibnamefont
  {{Blatt}}},\ }\href {\doibase 10.1088/1367-2630/13/7/073043} {\bibfield
  {journal} {\bibinfo  {journal} {New Journal of Physics}\ }\textbf {\bibinfo
  {volume} {13}},\ \bibinfo {eid} {073043} (\bibinfo {year}
  {2011})}\BibitemShut {NoStop}%
\bibitem [{\citenamefont {{Sterling}}\ \emph {et~al.}(2014)\citenamefont
  {{Sterling}}, \citenamefont {{Rattanasonti}}, \citenamefont {{Weidt}},
  \citenamefont {{Lake}}, \citenamefont {{Srinivasan}}, \citenamefont
  {{Webster}}, \citenamefont {{Kraft}},\ and\ \citenamefont
  {{Hensinger}}}]{Hensinger14natcom}%
  \BibitemOpen
  \bibfield  {author} {\bibinfo {author} {\bibfnamefont {R.~C.}\ \bibnamefont
  {{Sterling}}}, \bibinfo {author} {\bibfnamefont {H.}~\bibnamefont
  {{Rattanasonti}}}, \bibinfo {author} {\bibfnamefont {S.}~\bibnamefont
  {{Weidt}}}, \bibinfo {author} {\bibfnamefont {K.}~\bibnamefont {{Lake}}},
  \bibinfo {author} {\bibfnamefont {P.}~\bibnamefont {{Srinivasan}}}, \bibinfo
  {author} {\bibfnamefont {S.~C.}\ \bibnamefont {{Webster}}}, \bibinfo {author}
  {\bibfnamefont {M.}~\bibnamefont {{Kraft}}}, \ and\ \bibinfo {author}
  {\bibfnamefont {W.~K.}\ \bibnamefont {{Hensinger}}},\ }\href {\doibase
  10.1038/ncomms4637} {\bibfield  {journal} {\bibinfo  {journal} {Nature
  Communications}\ }\textbf {\bibinfo {volume} {5}},\ \bibinfo {eid} {3637}
  (\bibinfo {year} {2014})}\BibitemShut {NoStop}%
\bibitem [{\citenamefont {Marcos}\ \emph {et~al.}(2013)\citenamefont {Marcos},
  \citenamefont {Rabl}, \citenamefont {Rico},\ and\ \citenamefont
  {Zoller}}]{Marcos13prl}%
  \BibitemOpen
  \bibfield  {author} {\bibinfo {author} {\bibfnamefont {D.}~\bibnamefont
  {Marcos}}, \bibinfo {author} {\bibfnamefont {P.}~\bibnamefont {Rabl}},
  \bibinfo {author} {\bibfnamefont {E.}~\bibnamefont {Rico}}, \ and\ \bibinfo
  {author} {\bibfnamefont {P.}~\bibnamefont {Zoller}},\ }\href {\doibase
  10.1103/PhysRevLett.111.110504} {\bibfield  {journal} {\bibinfo  {journal}
  {Phys. Rev. Lett.}\ }\textbf {\bibinfo {volume} {111}},\ \bibinfo {pages}
  {110504} (\bibinfo {year} {2013})}\BibitemShut {NoStop}%
\bibitem [{\citenamefont {Hauke}\ \emph {et~al.}(2013)\citenamefont {Hauke},
  \citenamefont {Marcos}, \citenamefont {Dalmonte},\ and\ \citenamefont
  {Zoller}}]{Hauke13prx}%
  \BibitemOpen
  \bibfield  {author} {\bibinfo {author} {\bibfnamefont {P.}~\bibnamefont
  {Hauke}}, \bibinfo {author} {\bibfnamefont {D.}~\bibnamefont {Marcos}},
  \bibinfo {author} {\bibfnamefont {M.}~\bibnamefont {Dalmonte}}, \ and\
  \bibinfo {author} {\bibfnamefont {P.}~\bibnamefont {Zoller}},\ }\href
  {\doibase 10.1103/PhysRevX.3.041018} {\bibfield  {journal} {\bibinfo
  {journal} {Phys. Rev. X}\ }\textbf {\bibinfo {volume} {3}},\ \bibinfo {pages}
  {041018} (\bibinfo {year} {2013})}\BibitemShut {NoStop}%
\bibitem [{\citenamefont {Kogut}(1979)}]{kogut79rmp}%
  \BibitemOpen
  \bibfield  {author} {\bibinfo {author} {\bibfnamefont {J.~B.}\ \bibnamefont
  {Kogut}},\ }\href {\doibase 10.1103/RevModPhys.51.659} {\bibfield  {journal}
  {\bibinfo  {journal} {Rev. Mod. Phys.}\ }\textbf {\bibinfo {volume} {51}},\
  \bibinfo {pages} {659} (\bibinfo {year} {1979})}\BibitemShut {NoStop}%
\bibitem [{\citenamefont {Suzuki}\ \emph {et~al.}(2013)\citenamefont {Suzuki},
  \citenamefont {Inoue},\ and\ \citenamefont {Chakrabarti}}]{ChakrabartiBook}%
  \BibitemOpen
  \bibfield  {author} {\bibinfo {author} {\bibfnamefont {S.}~\bibnamefont
  {Suzuki}}, \bibinfo {author} {\bibfnamefont {J.}~\bibnamefont {Inoue}}, \
  and\ \bibinfo {author} {\bibfnamefont {B.~K.}\ \bibnamefont {Chakrabarti}},\
  }\href@noop {} {\emph {\bibinfo {title} {Quantum Ising Phases and Transitions
  in Transverse Ising Models}}}\ (\bibinfo  {publisher} {Springer},\ \bibinfo
  {year} {2013})\BibitemShut {NoStop}%
\bibitem [{\citenamefont {{Ivanov}}\ and\ \citenamefont
  {{Porras}}(2013)}]{Ivanov13pra}%
  \BibitemOpen
  \bibfield  {author} {\bibinfo {author} {\bibfnamefont {P.~A.}\ \bibnamefont
  {{Ivanov}}}\ and\ \bibinfo {author} {\bibfnamefont {D.}~\bibnamefont
  {{Porras}}},\ }\href {\doibase 10.1103/PhysRevA.88.023803} {\bibfield
  {journal} {\bibinfo  {journal} {\pra}\ }\textbf {\bibinfo {volume} {88}},\
  \bibinfo {eid} {023803} (\bibinfo {year} {2013})}\BibitemShut {NoStop}%
\bibitem [{\citenamefont {Porras}\ and\ \citenamefont
  {Cirac}(2004)}]{Porras04prl}%
  \BibitemOpen
  \bibfield  {author} {\bibinfo {author} {\bibfnamefont {D.}~\bibnamefont
  {Porras}}\ and\ \bibinfo {author} {\bibfnamefont {J.}~\bibnamefont {Cirac}},\
  }\href {\doibase 10.1103/PhysRevLett.92.207901} {\bibfield  {journal}
  {\bibinfo  {journal} {Phys. Rev. Lett.}\ }\textbf {\bibinfo {volume} {92}},\
  \bibinfo {pages} {207901} (\bibinfo {year} {2004})}\BibitemShut {NoStop}%
\bibitem [{\citenamefont {Baym}(1969)}]{BaymBook}%
  \BibitemOpen
  \bibfield  {author} {\bibinfo {author} {\bibfnamefont {G.}~\bibnamefont
  {Baym}},\ }\href@noop {} {\emph {\bibinfo {title} {Lectures on Quantum
  Mechanics}}}\ (\bibinfo  {publisher} {Benjamin},\ \bibinfo {year}
  {1969})\BibitemShut {NoStop}%
\bibitem [{\citenamefont {Silbey}\ and\ \citenamefont
  {Harris}(1984)}]{Silbey84jourchemphys}%
  \BibitemOpen
  \bibfield  {author} {\bibinfo {author} {\bibfnamefont {R.}~\bibnamefont
  {Silbey}}\ and\ \bibinfo {author} {\bibfnamefont {R.~A.}\ \bibnamefont
  {Harris}},\ }\href {\doibase http://dx.doi.org/10.1063/1.447055} {\bibfield
  {journal} {\bibinfo  {journal} {The Journal of Chemical Physics}\ }\textbf
  {\bibinfo {volume} {80}},\ \bibinfo {pages} {2615} (\bibinfo {year}
  {1984})}\BibitemShut {NoStop}%
\bibitem [{\citenamefont {{Kurcz}}\ \emph {et~al.}(2014)\citenamefont
  {{Kurcz}}, \citenamefont {{Garcia-Ripoll}},\ and\ \citenamefont
  {{Bermudez}}}]{Kurz14arX}%
  \BibitemOpen
  \bibfield  {author} {\bibinfo {author} {\bibfnamefont {A.}~\bibnamefont
  {{Kurcz}}}, \bibinfo {author} {\bibfnamefont {J.~J.}\ \bibnamefont
  {{Garcia-Ripoll}}}, \ and\ \bibinfo {author} {\bibfnamefont {A.}~\bibnamefont
  {{Bermudez}}},\ }\href@noop {} {\bibfield  {journal} {\bibinfo  {journal}
  {ArXiv e-prints}\ } (\bibinfo {year} {2014})},\ \Eprint
  {http://arxiv.org/abs/1408.1878} {arXiv:1408.1878 [quant-ph]} \BibitemShut
  {NoStop}%
\bibitem [{\citenamefont {White}(1993)}]{White93prb}%
  \BibitemOpen
  \bibfield  {author} {\bibinfo {author} {\bibfnamefont {S.~R.}\ \bibnamefont
  {White}},\ }\href {\doibase 10.1103/PhysRevB.48.10345} {\bibfield  {journal}
  {\bibinfo  {journal} {Phys. Rev. B}\ }\textbf {\bibinfo {volume} {48}},\
  \bibinfo {pages} {10345} (\bibinfo {year} {1993})}\BibitemShut {NoStop}%
\bibitem [{\citenamefont {Schollw{\"o}ck}(2011)}]{Schollwoeck11aop}%
  \BibitemOpen
  \bibfield  {author} {\bibinfo {author} {\bibfnamefont {U.}~\bibnamefont
  {Schollw{\"o}ck}},\ }\href {\doibase
  http://dx.doi.org/10.1016/j.aop.2010.09.012} {\bibfield  {journal} {\bibinfo
  {journal} {Annals of Physics}\ }\textbf {\bibinfo {volume} {326}},\ \bibinfo
  {pages} {96 } (\bibinfo {year} {2011})}\BibitemShut {NoStop}%
\bibitem [{\citenamefont {Sachdev}(2011)}]{SachdevBook}%
  \BibitemOpen
  \bibfield  {author} {\bibinfo {author} {\bibfnamefont {S.}~\bibnamefont
  {Sachdev}},\ }\href@noop {} {\emph {\bibinfo {title} {Quantum Phase
  Transitions}}}\ (\bibinfo  {publisher} {Cambridge University Press},\
  \bibinfo {year} {2011})\BibitemShut {NoStop}%
\bibitem [{\citenamefont {Leibfried}\ \emph {et~al.}(2003)\citenamefont
  {Leibfried}, \citenamefont {Blatt}, \citenamefont {Monroe},\ and\
  \citenamefont {Wineland}}]{Leibfried03rmp}%
  \BibitemOpen
  \bibfield  {author} {\bibinfo {author} {\bibfnamefont {D.}~\bibnamefont
  {Leibfried}}, \bibinfo {author} {\bibfnamefont {R.}~\bibnamefont {Blatt}},
  \bibinfo {author} {\bibfnamefont {C.}~\bibnamefont {Monroe}}, \ and\ \bibinfo
  {author} {\bibfnamefont {D.}~\bibnamefont {Wineland}},\ }\href {\doibase
  10.1103/RevModPhys.75.281} {\bibfield  {journal} {\bibinfo  {journal} {Rev.
  Mod. Phys.}\ }\textbf {\bibinfo {volume} {75}},\ \bibinfo {pages} {281}
  (\bibinfo {year} {2003})}\BibitemShut {NoStop}%
\bibitem [{\citenamefont {Saffman}\ \emph {et~al.}(2010)\citenamefont
  {Saffman}, \citenamefont {Walker},\ and\ \citenamefont
  {M\o{}lmer}}]{Molmer10rmp}%
  \BibitemOpen
  \bibfield  {author} {\bibinfo {author} {\bibfnamefont {M.}~\bibnamefont
  {Saffman}}, \bibinfo {author} {\bibfnamefont {T.~G.}\ \bibnamefont {Walker}},
  \ and\ \bibinfo {author} {\bibfnamefont {K.}~\bibnamefont {M\o{}lmer}},\
  }\href {\doibase 10.1103/RevModPhys.82.2313} {\bibfield  {journal} {\bibinfo
  {journal} {Rev. Mod. Phys.}\ }\textbf {\bibinfo {volume} {82}},\ \bibinfo
  {pages} {2313} (\bibinfo {year} {2010})}\BibitemShut {NoStop}%
\bibitem [{\citenamefont {{Deng}}\ \emph {et~al.}(2008)\citenamefont {{Deng}},
  \citenamefont {{Porras}},\ and\ \citenamefont {{Cirac}}}]{Deng08pra}%
  \BibitemOpen
  \bibfield  {author} {\bibinfo {author} {\bibfnamefont {X.-L.}\ \bibnamefont
  {{Deng}}}, \bibinfo {author} {\bibfnamefont {D.}~\bibnamefont {{Porras}}}, \
  and\ \bibinfo {author} {\bibfnamefont {J.~I.}\ \bibnamefont {{Cirac}}},\
  }\href {\doibase 10.1103/PhysRevA.77.033403} {\bibfield  {journal} {\bibinfo
  {journal} {\pra}\ }\textbf {\bibinfo {volume} {77}},\ \bibinfo {eid} {033403}
  (\bibinfo {year} {2008})},\ \Eprint {http://arxiv.org/abs/quant-ph/0703178}
  {quant-ph/0703178} \BibitemShut {NoStop}%
\bibitem [{\citenamefont {{Haze}}\ \emph {et~al.}(2012)\citenamefont {{Haze}},
  \citenamefont {{Tateishi}}, \citenamefont {{Noguchi}}, \citenamefont
  {{Toyoda}},\ and\ \citenamefont {{Urabe}}}]{Haze12pra}%
  \BibitemOpen
  \bibfield  {author} {\bibinfo {author} {\bibfnamefont {S.}~\bibnamefont
  {{Haze}}}, \bibinfo {author} {\bibfnamefont {Y.}~\bibnamefont {{Tateishi}}},
  \bibinfo {author} {\bibfnamefont {A.}~\bibnamefont {{Noguchi}}}, \bibinfo
  {author} {\bibfnamefont {K.}~\bibnamefont {{Toyoda}}}, \ and\ \bibinfo
  {author} {\bibfnamefont {S.}~\bibnamefont {{Urabe}}},\ }\href {\doibase
  10.1103/PhysRevA.85.031401} {\bibfield  {journal} {\bibinfo  {journal}
  {\pra}\ }\textbf {\bibinfo {volume} {85}},\ \bibinfo {eid} {031401} (\bibinfo
  {year} {2012})}\BibitemShut {NoStop}%
\bibitem [{\citenamefont {Wineland}\ \emph {et~al.}(2003)\citenamefont
  {Wineland}, \citenamefont {Barrett}, \citenamefont {Britton}, \citenamefont
  {Chiaverini}, \citenamefont {DeMarco}, \citenamefont {Itano}, \citenamefont
  {Jelenkovi{\'c}}, \citenamefont {Langer}, \citenamefont {Leibfried},
  \citenamefont {Meyer}, \citenamefont {Rosenband},\ and\ \citenamefont
  {Sch{\"a}tz}}]{Wineland03ptrs}%
  \BibitemOpen
  \bibfield  {author} {\bibinfo {author} {\bibfnamefont {D.~J.}\ \bibnamefont
  {Wineland}}, \bibinfo {author} {\bibfnamefont {M.}~\bibnamefont {Barrett}},
  \bibinfo {author} {\bibfnamefont {J.}~\bibnamefont {Britton}}, \bibinfo
  {author} {\bibfnamefont {J.}~\bibnamefont {Chiaverini}}, \bibinfo {author}
  {\bibfnamefont {B.}~\bibnamefont {DeMarco}}, \bibinfo {author} {\bibfnamefont
  {W.~M.}\ \bibnamefont {Itano}}, \bibinfo {author} {\bibfnamefont
  {B.}~\bibnamefont {Jelenkovi{\'c}}}, \bibinfo {author} {\bibfnamefont
  {C.}~\bibnamefont {Langer}}, \bibinfo {author} {\bibfnamefont
  {D.}~\bibnamefont {Leibfried}}, \bibinfo {author} {\bibfnamefont
  {V.}~\bibnamefont {Meyer}}, \bibinfo {author} {\bibfnamefont
  {T.}~\bibnamefont {Rosenband}}, \ and\ \bibinfo {author} {\bibfnamefont
  {T.}~\bibnamefont {Sch{\"a}tz}},\ }\href {\doibase 10.1098/rsta.2003.1205}
  {\bibfield  {journal} {\bibinfo  {journal} {Philosophical Transactions of the
  Royal Society of London A: Mathematical, Physical and Engineering Sciences}\
  }\textbf {\bibinfo {volume} {361}},\ \bibinfo {pages} {1349} (\bibinfo {year}
  {2003})}\BibitemShut {NoStop}%
\bibitem [{\citenamefont {Schiff}(1968)}]{SchiffBook}%
  \BibitemOpen
  \bibfield  {author} {\bibinfo {author} {\bibfnamefont {L.}~\bibnamefont
  {Schiff}},\ }\href@noop {} {\emph {\bibinfo {title} {Quantum Mechanics}}}\
  (\bibinfo  {publisher} {McGraw-Hill},\ \bibinfo {year} {1968})\BibitemShut
  {NoStop}%
\end{thebibliography}%

\end{document}